\documentclass[12pt,preprint]{aastex6}

\slugcomment{Revised version submitted to the Astronomical Journal 09 August 2016}
\begin{document}
\setcounter{figure}{0}
\title{Proper Motion of the Leo~II Dwarf Galaxy Based On\\
\textit{Hubble Space Telescope} Imaging.\footnote{Based on observations
with the NASA/ESA \textit{Hubble Space Telescope}, obtained at the Space
Telescope Science Institute, which is operated by the Association of
Universities for Research in Astronomy, Inc., under NASA contract NAS
5-26555.  The observations are associated with the program 11697.}
\footnote{Observations reported here were obtained at the MMT
Observatory, a joint facility of the University of Arizona and the
Smithsonian Institution.}}

\author{Slawomir Piatek} \affil{Dept. of Physics, New Jersey Institute
of Technology, Newark, NJ 07102}
\email{piatek@physics.rutgers.edu}

\author{Carlton Pryor}
\affil{Dept. of Physics and Astronomy, Rutgers, the State University
of New Jersey, 136~Frelinghuysen Rd., Piscataway, NJ 08854--8019}
\email{pryor@physics.rutgers.edu}

\author{Edward W.\ Olszewski}
\affil{Steward Observatory, The University of Arizona, 933 N.\ Cherry Avenue,
    Tucson, AZ 85721}
\email{eolszewski@as.arizona.edu}

\begin{abstract}

This article reports a measurement of the proper motion of Leo~II, a
dwarf galaxy that is a likely satellite of the Milky Way, based on
imaging with the Hubble Space Telescope and Wide Field Camera~3.  The
measurement uses compact background galaxies as standards of rest in
both channels of the camera for two distinct pointings of the telescope
and a QSO in one channel for each pointing, resulting in the weighted
average of six measurements.  The measured proper motion in the the
equatorial coordinate system is $(\mu_{\alpha},\mu_{\delta}) = (-6.9\pm
3.7, -8.7 \pm 3.9)$~mas~century$^{-1}$ and in the Galactic coordinate
system is $(\mu_{\ell},\mu_{b}) = (6.2\pm 3.9, -9.2 \pm
3.7)$~mas~century$^{-1}$. The implied space velocity with respect to the
Galactic center is $(\Pi, \Theta, Z) = (-37\pm 38, 117\pm 43, 40\pm
16)$~km~s$^{-1}$ or, expressed in Galactocentric radial and tangential
components, $(V_r, V_{tan}) = (21.9\pm 1.5, 127 \pm 42)$~km~s$^{-1}$.
The space velocity implies that the instantaneous orbital inclination is
$68^\circ$, with a 95\% confidence interval of $(66^\circ, 80^\circ)$. 
The measured motion supports the hypothesis that Leo~II, Leo~IV, Leo~V,
Crater~2, and the globular cluster Crater fell into the Milky Way as a
group.

\end{abstract}


\section{INTRODUCTION}
\label{intro}

        Among the original seven dwarf galaxies known in the halo of the
Milky Way, Leo~I and Leo~II are the most remote. \citet{bgr05}
estimates the heliocentric distance of Leo~II to be $233\pm15$~kpc from
the luminosity of the tip of the red giant branch in the $V-I$
color-magnitude diagram. Given its large distance, the question arises
whether Leo~II is a gravitationally-bound satellite of the Milky Way or
a galaxy just passing by.  Knowing the kinematics of the galaxy is
necessary to answer this question. \citet{k07} measured the systemic
heliocentric velocity to be $79.1\pm0.6$~km~s$^{-1}$ based on radial
velocities of 171 likely member stars, which is consistent with an
earlier measurement of $76.0\pm1.3$~km~s$^{-1}$ \citep{v95} based on a
sample of 31 stars within the core.  The radial velocity needs to be
augmented with a measurement of the proper motion to obtain the
instantaneous heliocentric space velocity of the dwarf. Transforming
this velocity into a coordinate system at rest at the center of the
Milky Way will, together with a knowledge of the Galactic potential,
determine whether Leo~II is bound.

      \citet{lkrk11} uses \textit{Hubble Space Telescope} (HST) imaging
with the Wide Field and Planetary Camera 2 and a time baseline of
14~years to make the only published measurement of the proper motion of
Leo~II: $(\mu_{\alpha}, \mu_{\delta})=(10.4\pm 11.3,
-3.3\pm 15.1)$~mas~century$^{-1}$ in the equatorial coordinate system. 
This result is derived from the motion of 3224 likely stars of Leo~II in
a coordinate system defined by 17 compact background galaxies.  The
study assumes a distance of 230~kpc and a radial velocity of
79~km~s$^{-1}$ to calculate, with additional assumptions about the Sun's
galactocentric distance and its motion with respect to the LSR,
galactocentric radial and tangential velocities of
$21.5\pm4.3$~km~s$^{-1}$ and $265.2\pm129.4$~km~s$^{-1}$, respectively. 
This velocity implies that the motion is mostly tangential; however, the
uncertainty in the tangential component is large compared to typical
Galactic velocities.  The current work contributes a measurement with
smaller uncertainty.

Deriving galactocentric motions requires adopting certain properties of
Leo~II and the Sun's galactocentric distance and motion with respect to
the LSR.   Table~\ref{tab:leoii} lists the values adopted in this article.

\floattable
\begin{deluxetable*}{lcc}
\tablecolumns{3}
\tablewidth{0pt} 
\tablecaption{Leo~II and the LSR at a Glance\label{tab:leoii}}
\tablehead{
\colhead{Quantity} &
\colhead{Value}    &
\colhead{Reference} \\
\colhead{(1)}&
\colhead{(2)}&
\colhead{(3)}}
\startdata
Right Ascension, $\alpha$ (J2000.0) & 11:13:28.8 & \citet{c07} \\
Declination, $\delta$ (J2000.0) & 22:09:06.0 & $^{\prime\prime}$ \\
Galactic longitude, $\ell$& $220\fdg 1691$& \\
Galactic latitude, $b$    & $+67\fdg 2313$& \\
Ellipticity, $e$ &$0.11$&  $^{\prime\prime}$  \\
Core radius, $r_{c}$&$2\farcm 64\pm 0\farcm 19$&$^{\prime\prime}$\\
Position angle, $\theta$ & $6\fdg 7\pm0\fdg 9$&$^{\prime\prime}$  \\
Heliocentric distance, $d$ & $233 \pm 15$~kpc & \citet{bgr05} \\
Heliocentric radial velocity, $v_{r}$&$79.1 \pm 0.6$~km~s$^{-1}$ & \citet{k07}\\
$R_{LSR}$&$8.2$~kpc&Bland-Hawthorn \& \\
$V_{LSR}$&$237$~km~s$^{-1}$& Gerhard (2016)\tablenotemark{a}\\
$u_{\odot}$ & $-10.0\pm 1.0$~km~s$^{-1}$ & $^{\prime\prime}$  \\
$v_{\odot}$ & $11.0\pm 2.0$~km~s$^{-1}$  & $^{\prime\prime}$ \\
$w_{\odot}$  & $7.0\pm 0.5$~km~s$^{-1}$ & $^{\prime\prime}$ \\
\enddata
\tablenotetext{a}{This article uses only $V_{LSR}+v_\odot$.
\citet{bhg16} argue that this combined motion is
known to an accuracy of $\pm 3$~km~s$^{-1}$.}
\end{deluxetable*}

The organization of the rest of the article is as follows.
Section~\ref{sec:data} describes the data, Section~\ref{sec:analysis}\
derives the proper motion, Section~\ref{sec:results}\ presents the
results and compares them to the previous measurement,
Section~\ref{sec:disc} discusses some implications of the proper motion,
and Section~\ref{sec:sum}\ briefly summarizes the results.
       
\section{OBSERVATIONS AND DATA}
\label{sec:data}

\begin{figure}[t!]
\centering
\includegraphics[angle=-90,scale=0.78]{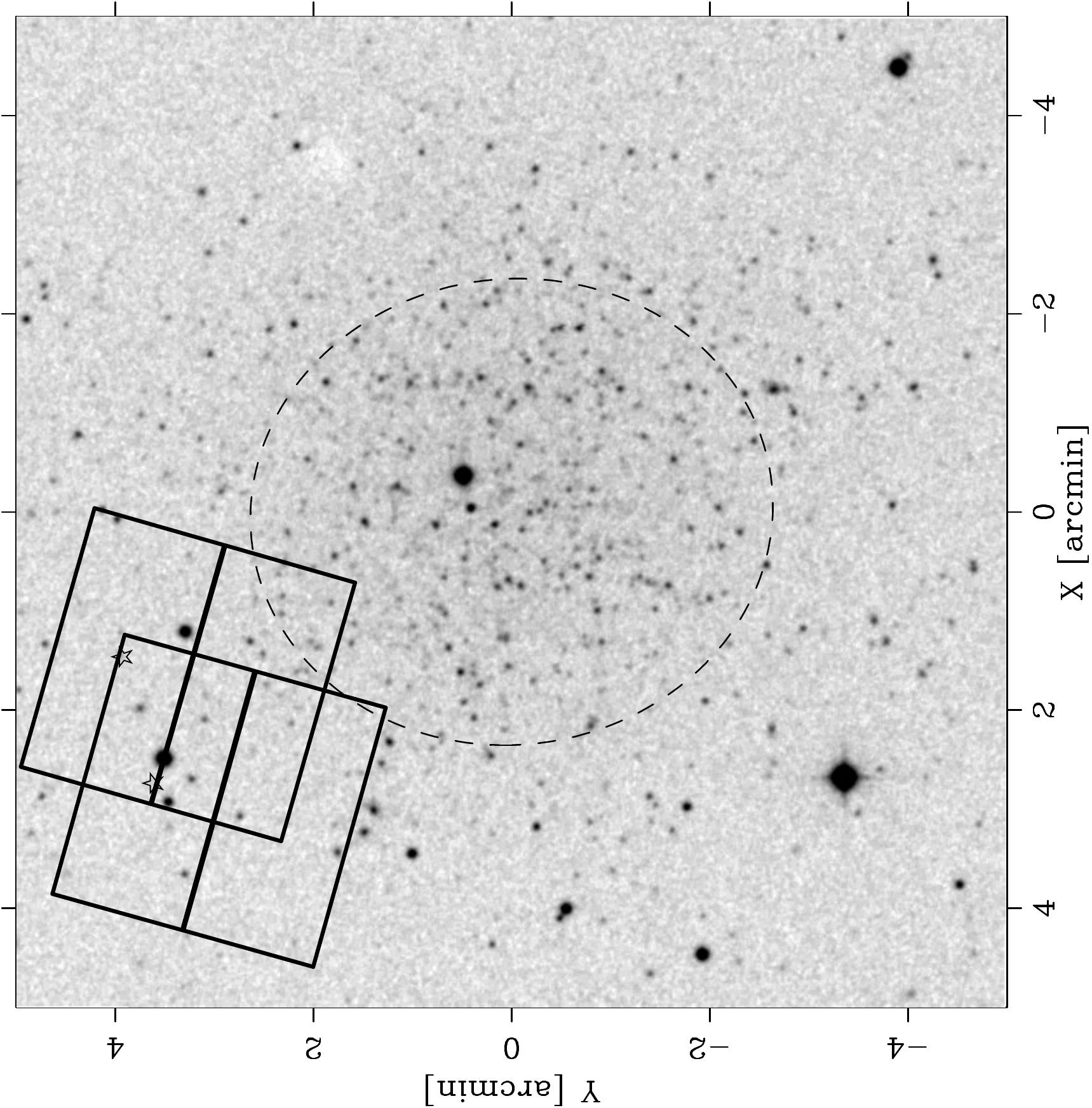}
\caption{Approximate location and orientation of the two fields in a
10$\times$10~arcmin$^{2}$ section of the sky centered on Leo~II.  North
is up and East is to the left.  The more northern and westward pointing
is Leo-II-1. The star symbol for each pointing represents the QSO in the
UVIS1 field. The dashed ellipse delineates the core of Leo-II. The
values for the center, position angle, ellipticity, and core radius of
the galaxy are from Table~\ref{tab:leoii}.  Note significant overlaps among the
fields.}
 \label{fig:leo2}
\end{figure}

        The data consist of HST imaging for two distinct pointings in
the direction of Leo~II taken at two epochs that are approximately two
years apart for both.  This article designates them as Leo-II-1 and
Leo-II-2.  The detector and filter are the Wide Field Camera~3 (WFC3)
and F350LP, respectively.  Table~\ref{tab:point} provides additional
details.  WFC3 has two 4096$\times$2051 CCDs or channels: UVIS1 and
UVIS2.  The direction of each pointing is that of a
spectroscopically-confirmed QSO located at or near the center of UVIS1. 
This work uses images from both the UVIS1 and UVIS2 fields; thus, there
are four distinct fields: Leo-II-1-UVIS1, Leo-II-1-UVIS2,
Leo-II-2-UVIS1, and Leo-II-2-UVIS2. Figure~\ref{fig:leo2} portrays the
locations of the fields in a 10$\times$10~arcmin$^{2}$ section of the
sky centered on Leo~II and taken from the STScI Digitized Sky
Survey\footnote{The Digitized Sky Surveys were produced at the Space
Telescope Science Institute under U.S. Government grant NAG W-2166. The
images of these surveys are based on photographic data obtained using
the Oschin Schmidt Telescope on Palomar Mountain and the UK Schmidt
Telescope. The plates were processed into the present compressed digital
form with the permission of these institutions.  The Second Palomar
Observatory Sky Survey (POSS-II) was made by the California Institute of
Technology with funds from the National Science Foundation, the National
Geographic Society, the Sloan Foundation, the Samuel Oschin Foundation,
and the Eastman Kodak Corporation.}.

\floattable
\begin{deluxetable*}{lclclll}
\tablecolumns{7}
\tablewidth{0pt} 
\tablecaption{Information about
Pointings and Images\label{tab:point}}
\tablehead{    &
\colhead{R.A.} &
\colhead{Decl.}&
\colhead{Date} &
               &
               &
\colhead{T$_{exp}$}\\
\colhead{Pointing}        &
\colhead{(J2000.0)}    &
\colhead{(J2000.0)}    &
\colhead{$yyyy-mm-dd$} &
\colhead{Detector}     &
\colhead{Filter}       &
\colhead{(s)}          \\
\colhead{(1)}&
\colhead{(2)}&
\colhead{(3)}&
\colhead{(4)}&
\colhead{(5)}&
\colhead{(6)}&
\colhead{(7)}}
\startdata
Leo~II-1&11:13:35&+22:13:02&$2010-05-12$&WFC3&F350LP&$4\times 633$\\
&&&&&&$8\times 661$ \\
&&&$2012-04-06$&&&$4 \times 626$\\
&&&&&&$8\times 653$ \\
\noalign{\vspace{3pt}}
Leo~II-2&11:13:41&+22:12:43&$2010-05-11$&WFC3&F350LP&$4 \times 633$\\
&&&&&&$8\times 661$\\
&&&$2012-05-01$&&&$4 \times 626$\\
&&&&&&$8\times 653 $ \\
\noalign{\vspace{3pt}}
\enddata
\end{deluxetable*}

\begin{figure}[t!]
\centering
\includegraphics[angle=-90,scale=0.79]{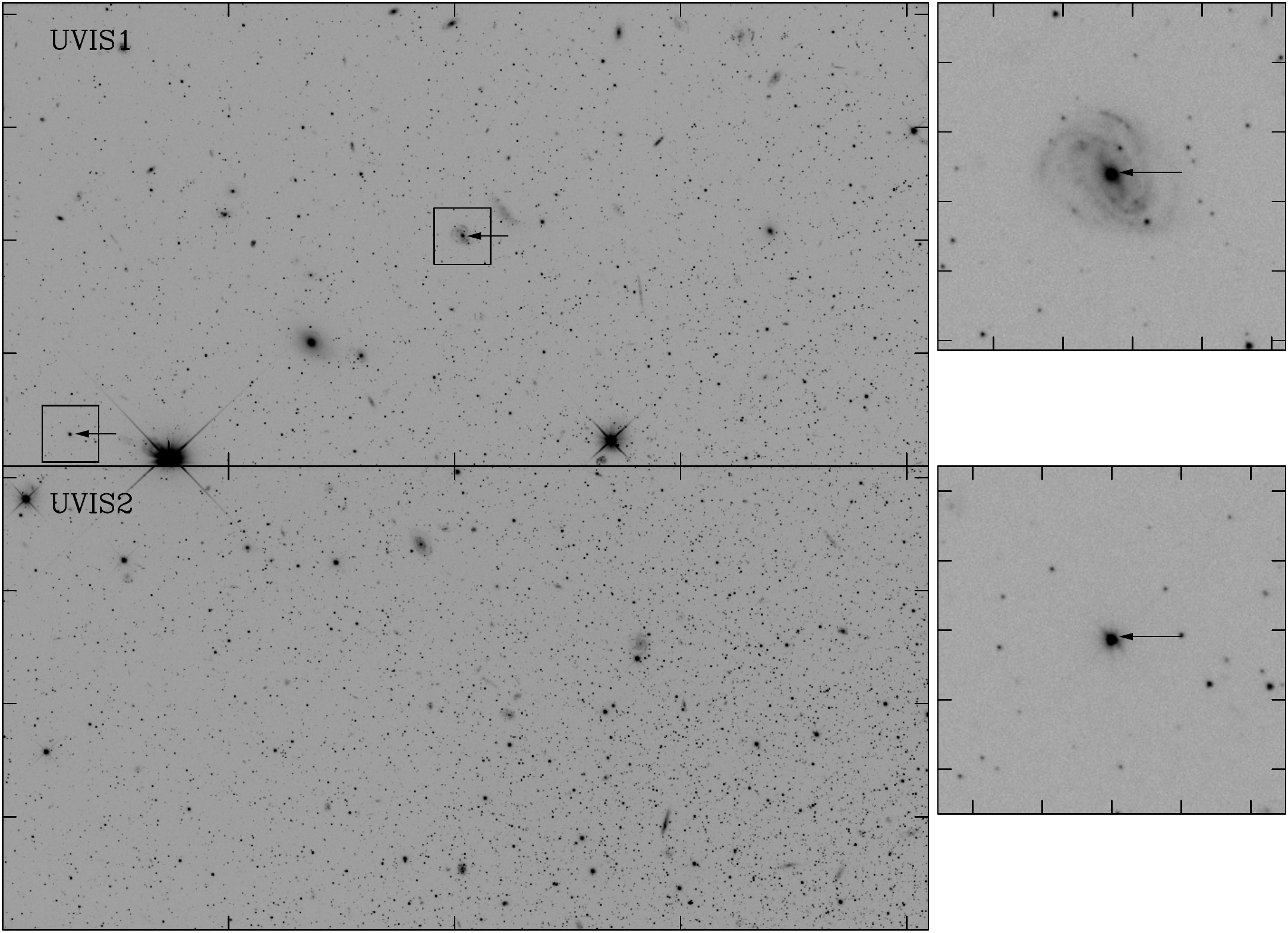}
\caption{The Leo-II-1-UVIS1 (top-left panel) and Leo-II-1-UVIS2
(bottom-left panel) fields. The QSO near the middle of the UVIS1 field
is at the center of a $250 \times 250$~pixel$^{2}$ box; an arrow points
to the QSO and the smaller top-right panel shows the section of the
image inside of this box.  Another QSO in the bottom-left region of the
UVIS1 field is also at the center of a $250 \times 250$~pixel$^{2}$ box.
An arrow points to this QSO and the bottom-right panel shows the section
of the image inside of this box.} \label{fig:leo21}
\end{figure}

\begin{figure}[t!]
\centering
\includegraphics[angle=-90,scale=0.79]{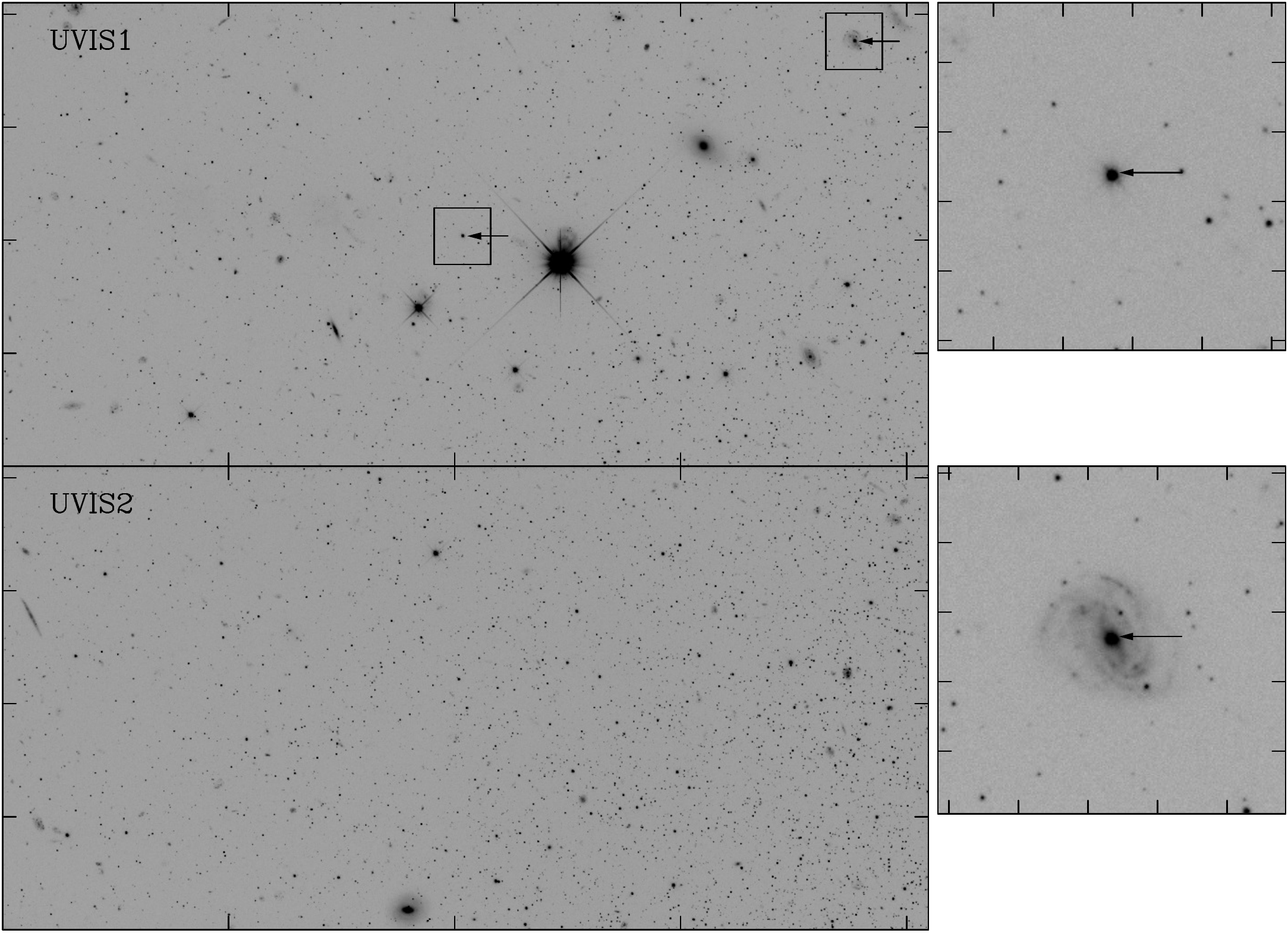}
\caption{Similar to Figure~\ref{fig:leo21} for the Leo-II-2-UVIS1 and
Leo-II-2-UVIS2 fields.}
\label{fig:leo22}
\end{figure}

Figure~\ref{fig:leo21} shows the Leo-II-1-UVIS1 (top left) and
Leo-II-1-UVIS2 (bottom left) fields.  Each shows the average, with
cosmic ray rejection, of 12 exposures taken at the first epoch. The QSO
near the center of the Leo-II-1-UVIS1 field is at the center of a
$250\times250$~pixel$^{2}$ box and an arrow points to it. The smaller
top-right panel depicts the region of the image within this box, with
the arrow again pointing to the QSO.  Because of the overlap between the
Leo-II-1-UVIS1 and Leo-II-2-UVIS1 fields, the QSO centered in the
Leo-II-2 pointing is also present in the lower left section of
Leo-II-1-UVIS1 and an arrow also points to it.  A $250 \times
250$~pixel$^{2}$ box is also centered on the QSO and the bottom right
panel shows the region of the image within this box, with an arrow again
pointing at the QSO.  Figure~\ref{fig:leo22} is analogous to
Fig.~\ref{fig:leo21} for the Leo-II-2 pointing.

\begin{figure}[p]
\centering
\includegraphics[angle=0,scale=0.85]{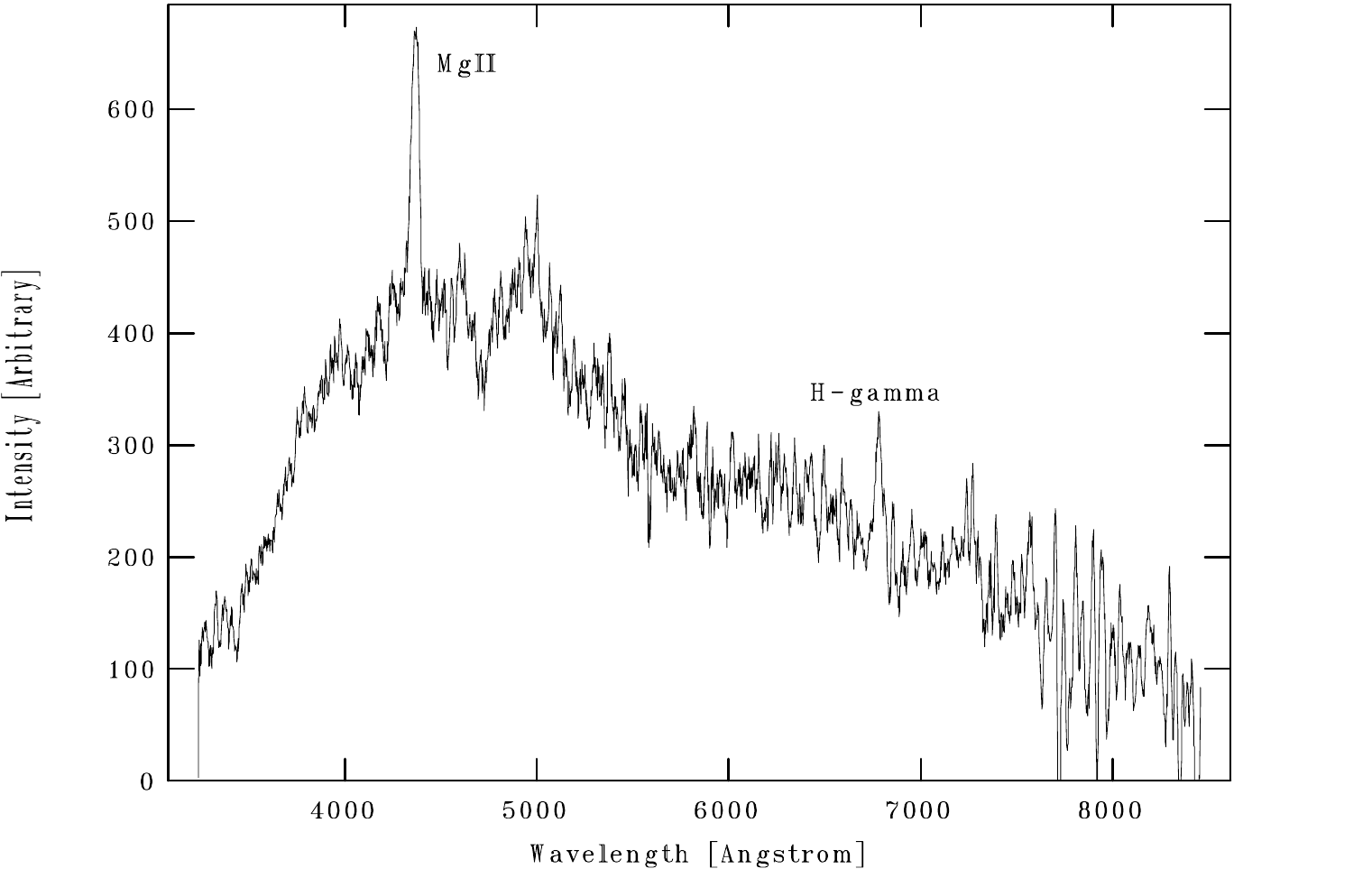}
\caption{Spectrum of Q1113+2213 in the Leo-II-1-UVIS1 field taken with
the Blue Channel Spectrograph and the 300-line grating on the MMT on
2008 January 15.  The spectrum covers 5200~\AA, the resolution is
22~\AA\ with the 3.5-arcsec slit, and the exposure time was 1800~s.  The
spectrum has been smoothed by a running mean of 9 points, which is
approximately the resolution. The measured redshift is 0.56.}
\label{fig:qso1}
\end{figure}

\begin{figure}[p]
\centering
\includegraphics[angle=0,scale=0.85]{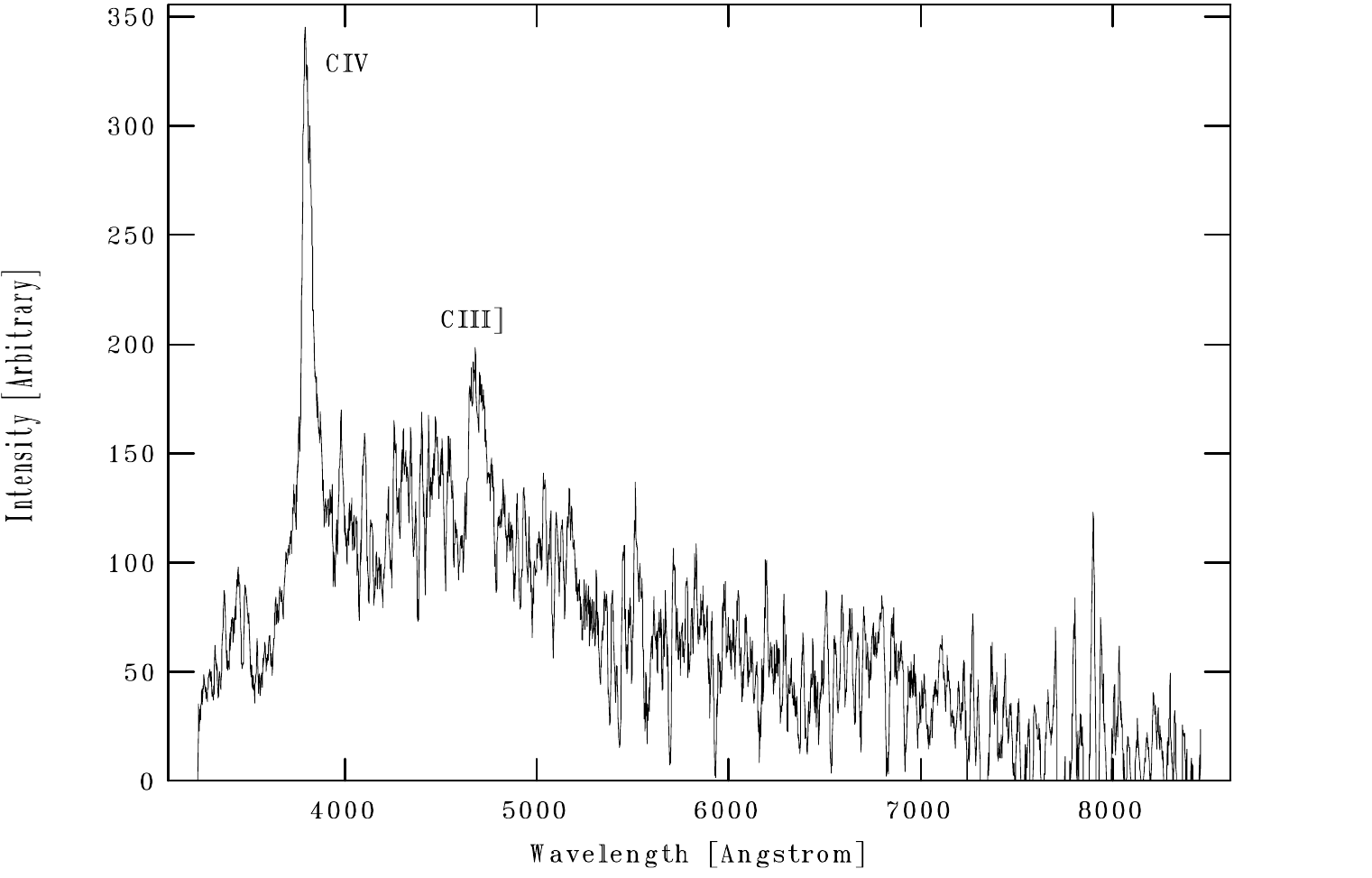}
\caption{Same as Fig.~\ref{fig:qso1}\ for Q1113+2212 in the
Leo-II-2-UVIS1 field. The exposure time was 1500~s and the measured
redshift is 1.45.}
\label{fig:qso2}
\end{figure}

        We identified the two QSOs with photometry from the Sloan Digital
Sky Survey Data Release 3 \citep{dr3} using the method described in
\citet{r05} and confirmed them with spectra taken with the MMT
Blue Channel Spectrograph on 2008 January 15.  Q1113+2213, at the center
of the Leo-II-1-UVIS1 field, has V=20.6 and the spectrum in
Figure~\ref{fig:qso1} implies a redshift of 0.56.
Q1113+2212, at the center of the Leo-II-2-UVIS1 field, has V=21.5 and
Figure~\ref{fig:qso2} implies a redshift of 1.45.  

        Measuring a proper motion involves taking images at different
epochs. If the images are taken with a CCD, the degradation with time of
the charge transfer efficiency (CTE) can introduce spurious motions, as
shown for the STIS detector by \citet{b04} and further discussed by
\citet{bpp05}.  A degrading CTE is present in all of the CCDs, past and
present, on HST.  Minimizing its effect on precision astrometry has been
a challenge and several techniques have been employed. 1) Numerical
modeling of the charge transfer process and correcting the existing
science images \citep{ba02}. 2) Multi-field imaging of the target with
different orientations of the CCD; thus, averaging out the effects of
degrading CTE \citep[\textit{e.g.},][]{kma06}. 3) Adopting an empirical
model for how degrading CTE affects the centroid of an object as a
function of its S/N and including the amplitude of this spurious shift
in the fit of the transformation between the epochs
\citep[\textit{e.g.},][]{ppo10}. 4) Pre-flashing the CCD, or introducing
a ``fat zero,'' to fill charge traps just before the science image is
taken \citep[\textit{e.g.},][]{ambn12}.  STScI now recommends this
approach for WFC3 if the average sky value is less than 12~electrons
\citep{d15}.  5) Injection of charge into every n-th row just before the
science image is taken \citep{bbgnp11}.  The injected charge fills
charge traps when injected and during the readout. The current study
adopts this approach --- the second-epoch images have injected charge
--- because it eliminates the cause of CTE at its source.  In
the Leo-II-1-UVIS1 field, injected charge is present in every
17$^{\mbox{th}}$ row starting with the 25$^{\mbox{th}}$ and ending with
the 2048$^{\mbox{th}}$.  This is designated as the (17, 25, 2048)
injection pattern.  The patterns for the remaining fields are (17, 4,
2027) for Leo-II-1-UVIS2, (17, 25, 2048) for Leo-II-2-UVIS1, and (17, 4,
2027) for Leo-II-2-UVIS2.  The actual amount of charge arriving at the
destination pixel depends on the history of charge loss and, possibly,
charge gain as the charge packet is clocked through the CCD.  The noise
in the charge injection is less than expected from Poisson statistics,
amounting to about 18~electrons per pixel in the injection row
\citep{bbgnp11}.

Because the first epoch data were taken soon after the installation of
the WFC3, when the CTE was high, we chose not to use the charge
injection option on these data, avoiding the added noise.  The wide
bandpass was also expected to produce significant average sky levels
(they are 56-80 electrons), reducing the effect of CTE degredation.  The
end of Section~\ref{sec:mpm} presents evidence that CTE losses in the
first-epoch data have not had a significant impact on our results.

\section{DATA REDUCTION AND ANALYSIS}
\label{sec:analysis}

        Measuring a proper motion of a resolved stellar system such as
Leo~II from HST imaging involves two major tasks. The first is
determining the most accurate centroids of stars and of bright and
compact features in background galaxies.  We visually inspected each
galaxy.  The second is matching the objects in the images at multiple
epochs and deriving the relative shift between the member stars and the
background galaxies and QSOs. The proper motion is proportional to this
shift. The methods for both steps employed in this work are nearly
identical to those in \citet{ppo15}; thus, the current work only briefly
comments on the minor differences, although it still presents the key
diagnostic plots.

\subsection{Measuring the Coordinates}
\label{sec:mcoord}

        The method for measuring the centroids of stars and galaxies
employed in this work differs from that described in \citet{ppo15} in
two ways.  The first is that the correction for the geometrical
distortions for the WFC3 with the F350LP filter come from \citet{kp14}.
This correction is more uncertain than those for filters used more
widely, but this is automatically accounted for by the empirical
estimates of our measurement uncertainties described in
Sections~\ref{sec:mpm} and \ref{sec:results}.

The second difference is the way that the images were corrected for the
degrading CTE.  The current work uses the charge injection technique,
described in Section~\ref{sec:data}.  The standard HST pipeline uses
charge-injected bias images to remove the injected charge, but the
removal is incomplete.  Therefore, before processing the images with
DOLPHOT \citep{dol00} to obtain the first estimate for a centroid and
flux of an object, our method subtracts from each row with charge
injection the difference between the robust (biweight) mean of the pixel
values in the row and the robust mean of the pixel values in six nearby
rows.  The uncertainty of each pixel value in a row with charge
injection is the sum in quadrature of a robust estimate of the standard
deviation of the values in the row and the Poisson noise of the net
signal in that pixel (the signal minus the robust row mean).

       At the end of identifying objects and measuring their positions,
there are 1153 objects common to the two epochs in the Leo-II-1-UVIS1
field: the central QSO, 101 selected galaxies (including the QSO in the
corner), and 1051 stars with S/N greater than 10 at the first epoch.
These can be succinctly written as (1153,1,101,1051).  These numbers are
(2330,0,145,2185) for Leo-II-1-UVIS2; (897,1,142,754) for
Leo-II-2-UVIS1; and (1459,0,176,1283) for Leo-II-2-UVIS2.  The S/N above
is the average of the values from the individual images at an epoch. 
The positions of both QSOs and the galaxies are determined in the same
way with templates derived from the individual images.

\begin{figure}[t!]
\centering
\includegraphics[angle=-90,scale=0.74]{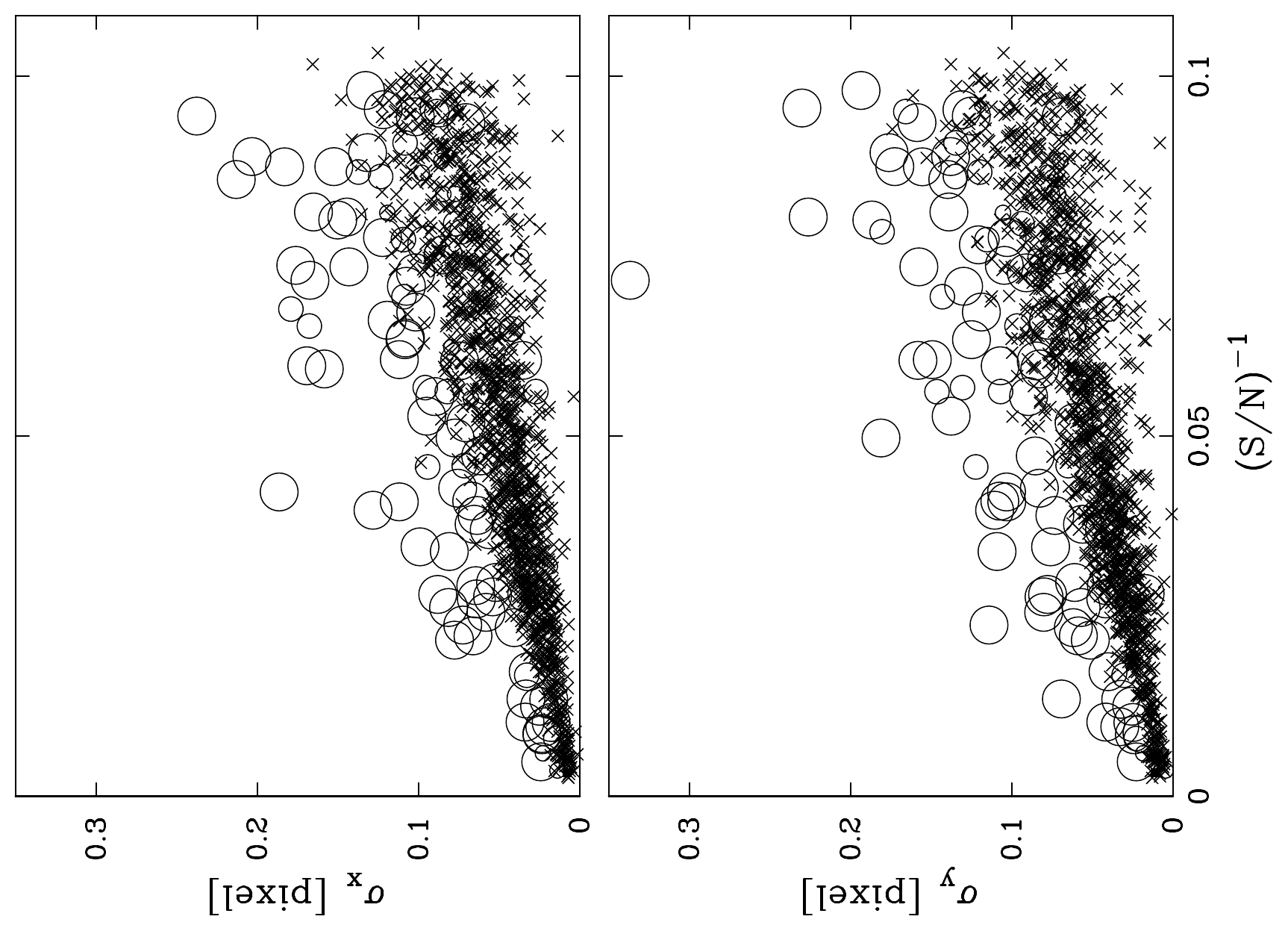}
\caption{The rms scatter around the mean of the X-component (top panel)
and the Y-component (bottom panel) of the centroid of an object as a
function of $(\mbox{S/N})^{-1}$ for the first-epoch exposures in the
Leo-II-1-UVIS1 field. The slanted cross is for a star and the open
circle for a galaxy, with the size of the circle proportional to the
FWHM of the galaxy image.}
\label{fig:rms}
\end{figure}

        Figure~\ref{fig:rms} shows the uncertainty in the location of
an object in the $X$ direction, $\sigma_x$, (top panel) and in the $Y$
direction, $\sigma_y$, (bottom panel) as a function of
$(\mbox{S/N})^{-1}$ for the Leo-II-1-UVIS1 field.  The estimate of the
uncertainty in a given direction is the scatter of the transformed
coordinates around their mean.  The figure shows a nearly linear
dependence of $\sigma_x$ and $\sigma_y$ on $(\mbox{S/N})^{-1}$, that
$\sigma_x \approx \sigma_y \approx 0.01$~pixel for the stars with the
highest S/N, and that $\sigma_x \approx \sigma_y \approx 0.05$~pixel for
S/N = 20.  The uncertainties for the other three fields are similar.

\subsection{Measuring the Proper Motion}
\label{sec:mpm}

           The standard coordinate system is a system co-moving with
Leo~II.  It is defined by the locations, corrected for geometrical
distortions, in the chronologically first exposure. Let $\mu_{x}$ and
$\mu_{y}$ be the components of the motion of an object, in
pixel~yr$^{-1}$, in this system.  The stars of Leo~II should have
$(\mu_{x},\mu_{y})$ consistent with zero, while the foreground stars and
background galaxies generally should not. The exception are those few
foreground stars that happen to have the same proper motion as Leo~II.
Contamination by Milky Way stars is unimportant in any case since the
Besan\c{c}on model \citep{robin03} predicts fewer than 10 such stars in
each of our fields. The identification of field stars is the same as in
our previous articles \citep[\textit{e.g.},][]{ppo15}: those whose
$\mu_x$ or $\mu_y$ differ from zero by a statistically significant
amount (have a contribution to the $\chi^2$ of the fit of the
transformation larger than 9.1, which corresponds approximately to a 1\%
chance of occurence). For the Leo~II data, eliminating all trends of
$\mu_x$ and $\mu_y$ with $X$ and $Y$ required the most general
second-order transformation plus a cubic term in $X$.  As in our
previous work \citep[\textit{e.g.},][]{ppo15}, it is necessary to add a
constant in quadrature to the coordinate uncertainties to make the
$\chi^2$ per degree of freedom of the fitted transformation equal to
one.  These values range from 0.0025 pixel to 0.0042 pixel, which are
similar to values found to be necessary previously.  The error bars in
all of the figures in this paper include these additive uncertainties.

\begin{figure}[p]
\centering
\includegraphics[angle=-90,scale=0.60]{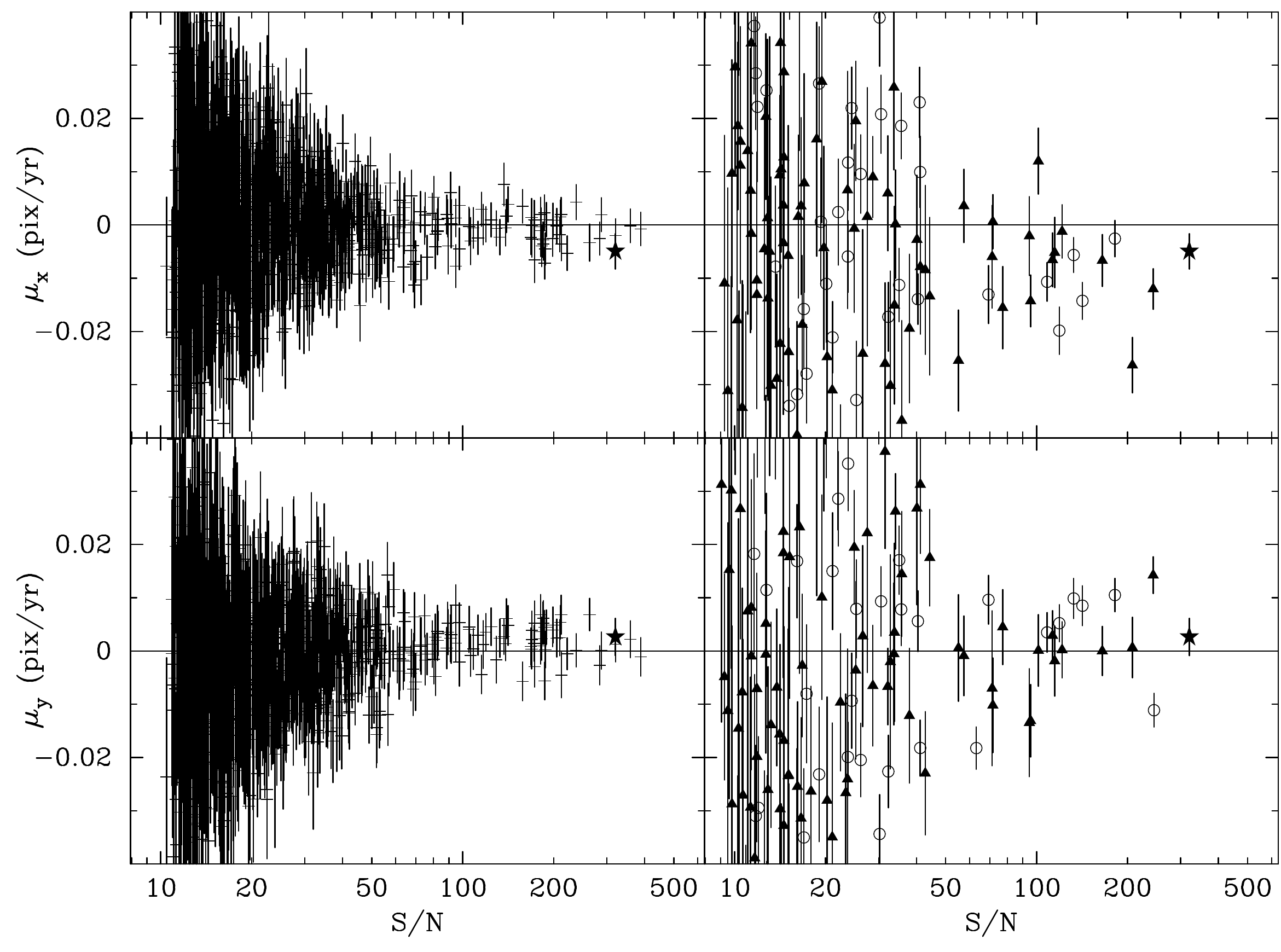}
\caption{Motion in the standard coordinate system, $\mu_{x}$ and
$\mu_{y}$, as a function of S/N for objects in the Leo-II-1-UVIS1 field.
The panels on the left are for the likely stars of Leo~II (plus signs)
and the QSO (filled star).  The panels on the right are for the
background galaxies (filled triangles), field stars (open circles), and,
again, the QSO.}
\label{fig:mu_sn_leo21_s}
\end{figure}

\begin{figure}[p]
\centering
\includegraphics[angle=-90,scale=0.60]{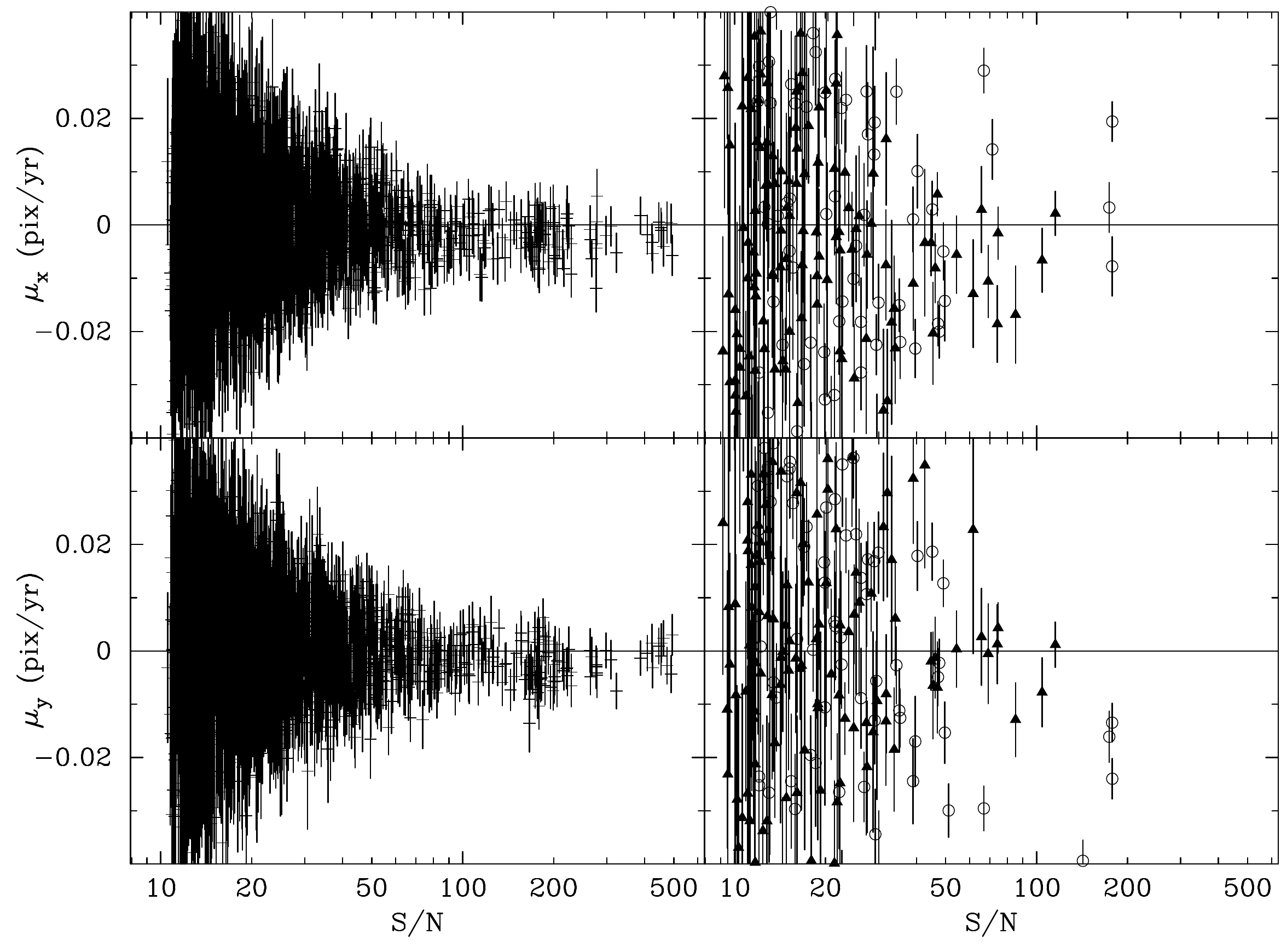}
\caption{The same as Figure~\ref{fig:mu_sn_leo21_s} for the Leo-II-1-UVIS2 field.}
\label{fig:mu_sn_leo21_t}
\end{figure}

\begin{figure}[p]
\centering
\includegraphics[angle=-90,scale=0.60]{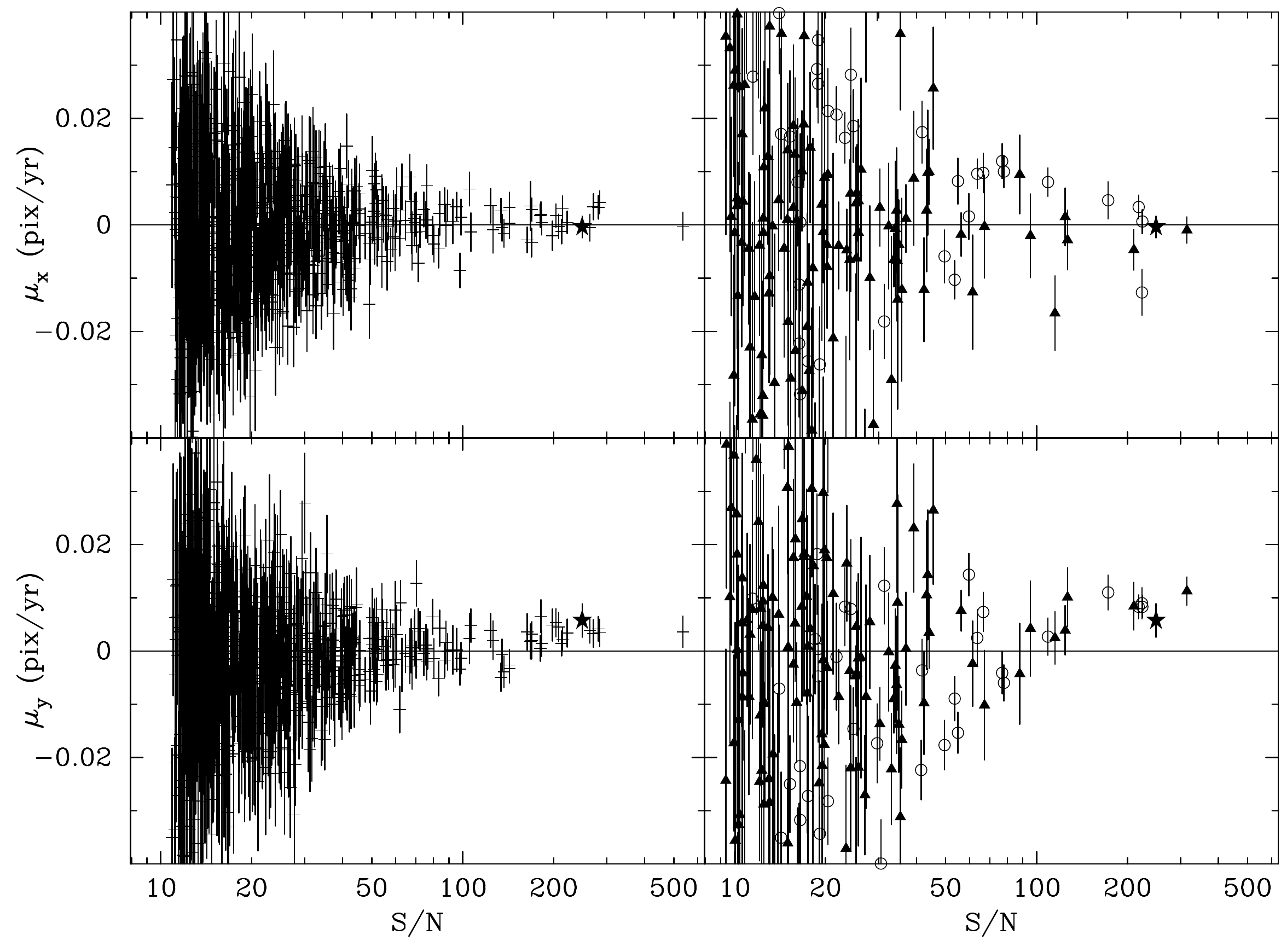}
\caption{The same as Figure~\ref{fig:mu_sn_leo21_s} for the Leo-II-2-UVIS1 field.}
\label{fig:mu_sn_leo22_s}
\end{figure}

\begin{figure}[p]
\centering
\includegraphics[angle=-90,scale=0.60]{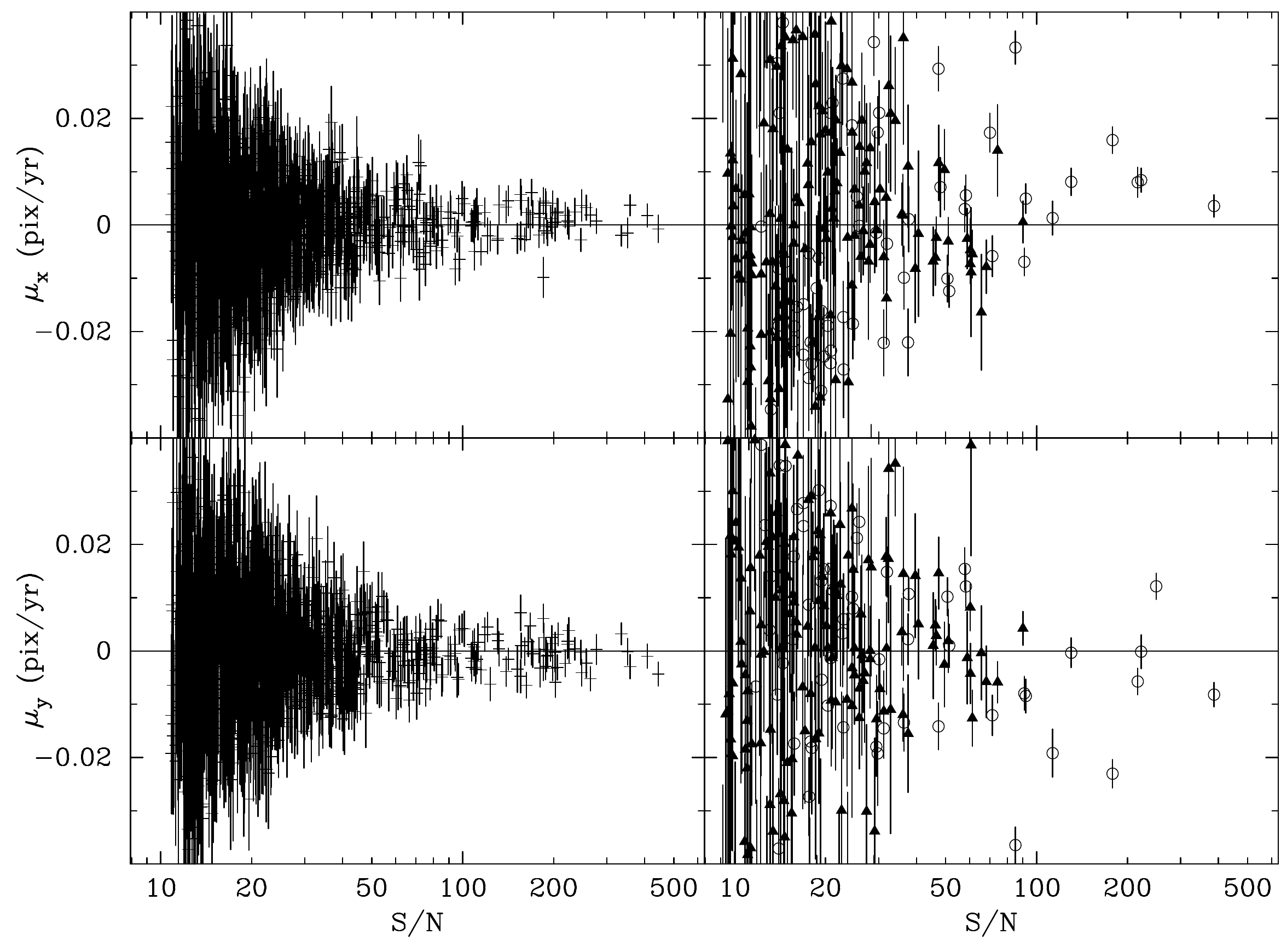}
\caption{The same as Figure~\ref{fig:mu_sn_leo21_s} for the Leo-II-2-UVIS2 field.}
\label{fig:mu_sn_leo22_t}
\end{figure}

         Figure~\ref{fig:mu_sn_leo21_s} shows $\mu_{x}$ (top panels)
and $\mu_{y}$ (bottom panels) as a function of S/N for the
Leo-II-1-UVIS1 field. The panels on the left are for the likely stars of
Leo~II (plus signs) and the QSO (a filled star), whereas the panels on
the right are for background galaxies (filled triangles), field stars
(open circles), and, again, the QSO.  Similarly,
Figures~\ref{fig:mu_sn_leo21_t}, \ref{fig:mu_sn_leo22_s}, and
\ref{fig:mu_sn_leo22_t} are for the other three fields. The left panels
in the four figures show that the mean motion $(<\mu_{x}>, <\mu_{y}>)$ for
the stars of Leo~II is consistent with zero.  As expected, the scatter
decreases with increasing S/N.  The panels on the right show that field
stars have a significant motion in the standard coordinate system. These
motions need not show any correlations or trends unless some of the
field stars are associated with each other. The motion of the background
galaxies should not show any trends with S/N; instead, the mean motion
should be offset from zero by an amount proportional to the proper
motion of Leo~II. The points for the galaxies have a larger scatter and
larger error bars than those for the stars and, thus, the offset from
zero is difficult to ascertain visually.

\begin{figure}[p]
\centering
\includegraphics[angle=-90,scale=0.60]{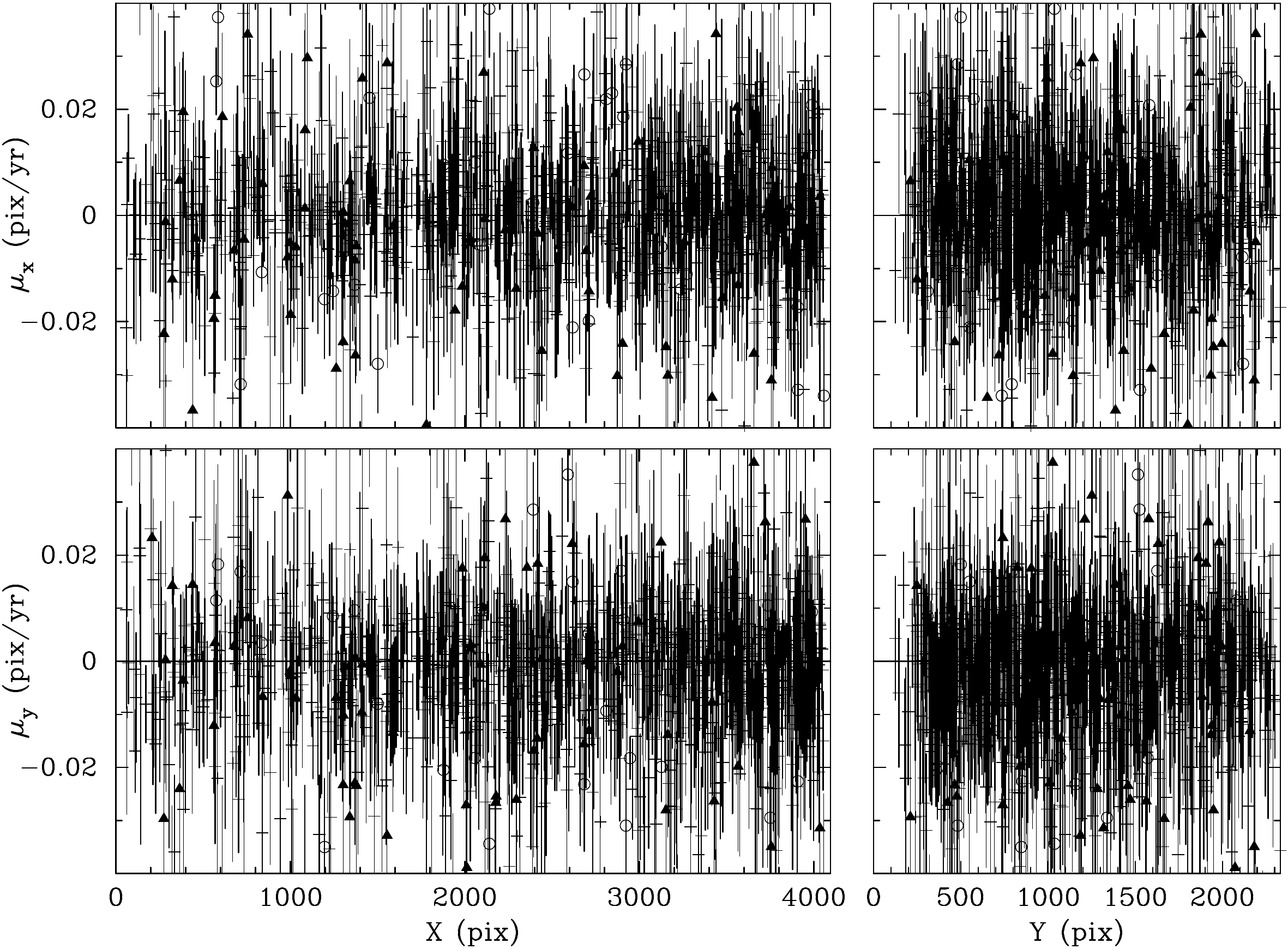}
\caption{Motions in the standard coordinate system, $\mu_{x}$ and
$\mu_{y}$, as a function of $X$ (left panels) and
$Y$ (right panels) for the Leo-II-1-UVIS1 field.  The symbols have the
same meaning as in the previous plot.}
\label{fig:mu_xy_leo21_s}
\end{figure}
\begin{figure}[p]
\centering
\includegraphics[angle=-90,scale=0.60]{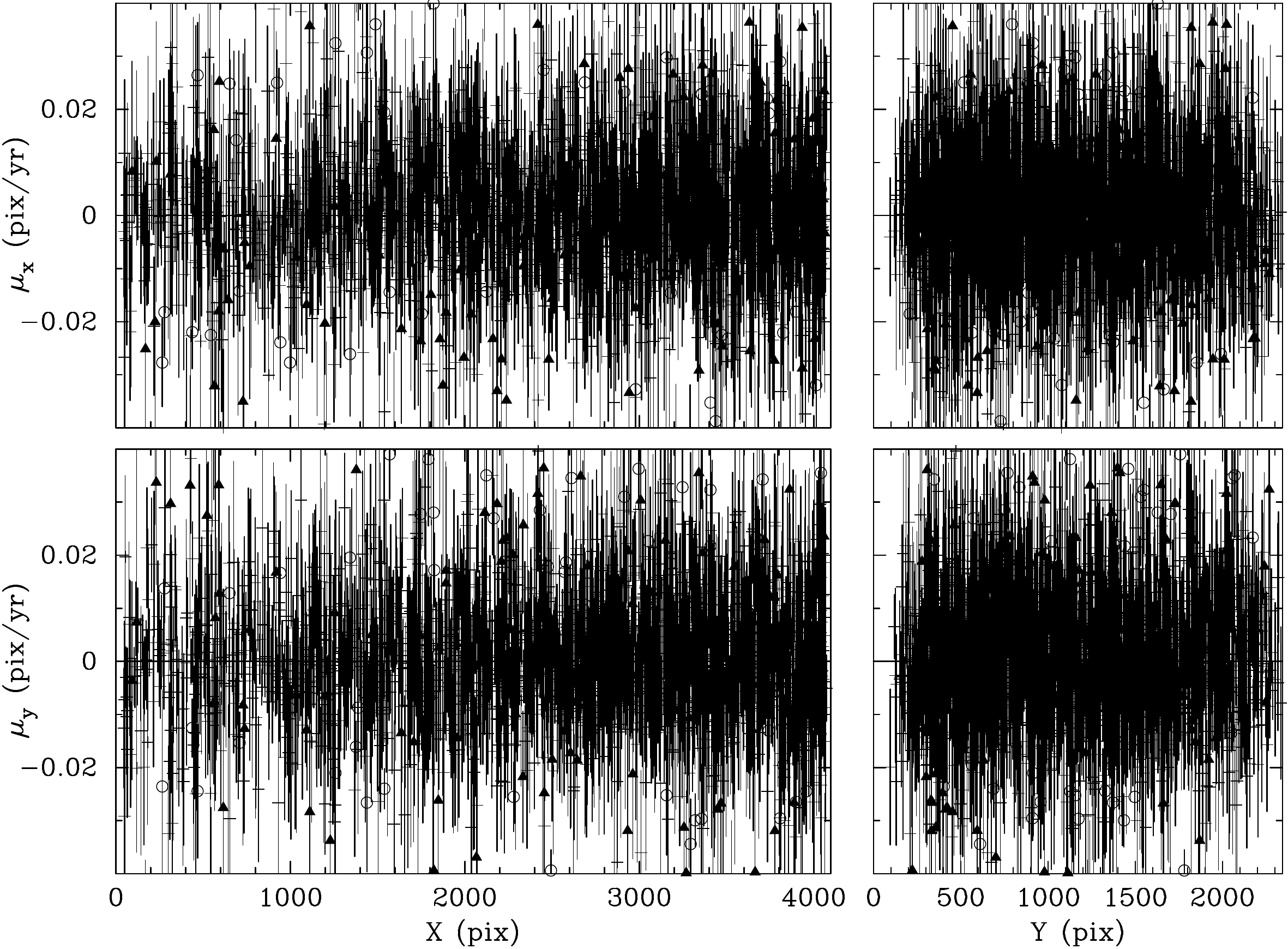}
\caption{The same as Figure~\ref{fig:mu_xy_leo21_s} for the Leo-II-1-UVIS2 field.}
\label{fig:mu_xy_leo21_t}
\end{figure}
\begin{figure}[p]
\centering
\includegraphics[angle=-90,scale=0.60]{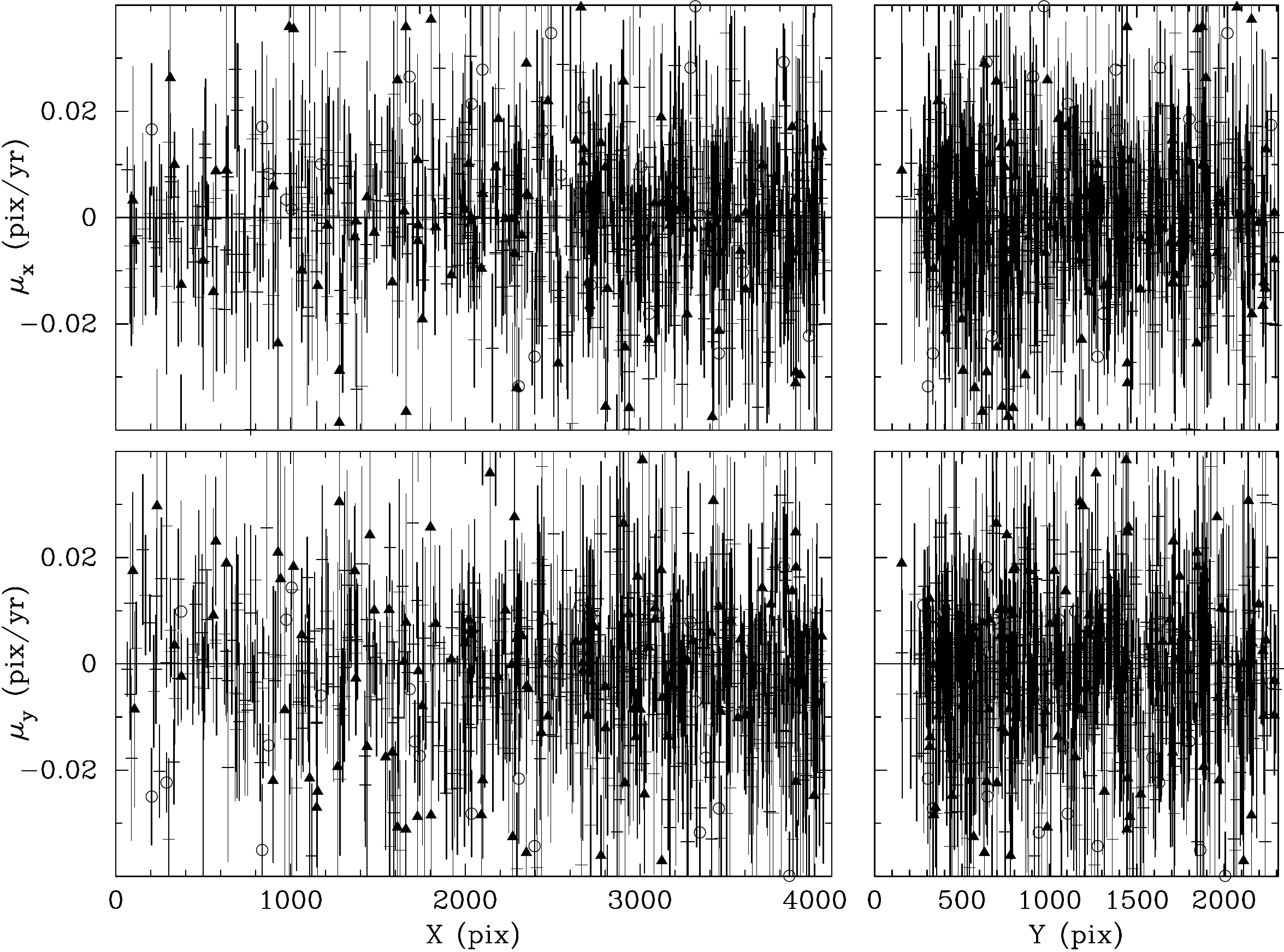}
\caption{The same as Figure~\ref{fig:mu_xy_leo21_s} for the Leo-II-2-UVIS1 field.}
\label{fig:mu_xy_leo22_s}
\end{figure}
\begin{figure}[p]
\centering
\includegraphics[angle=-90,scale=0.60]{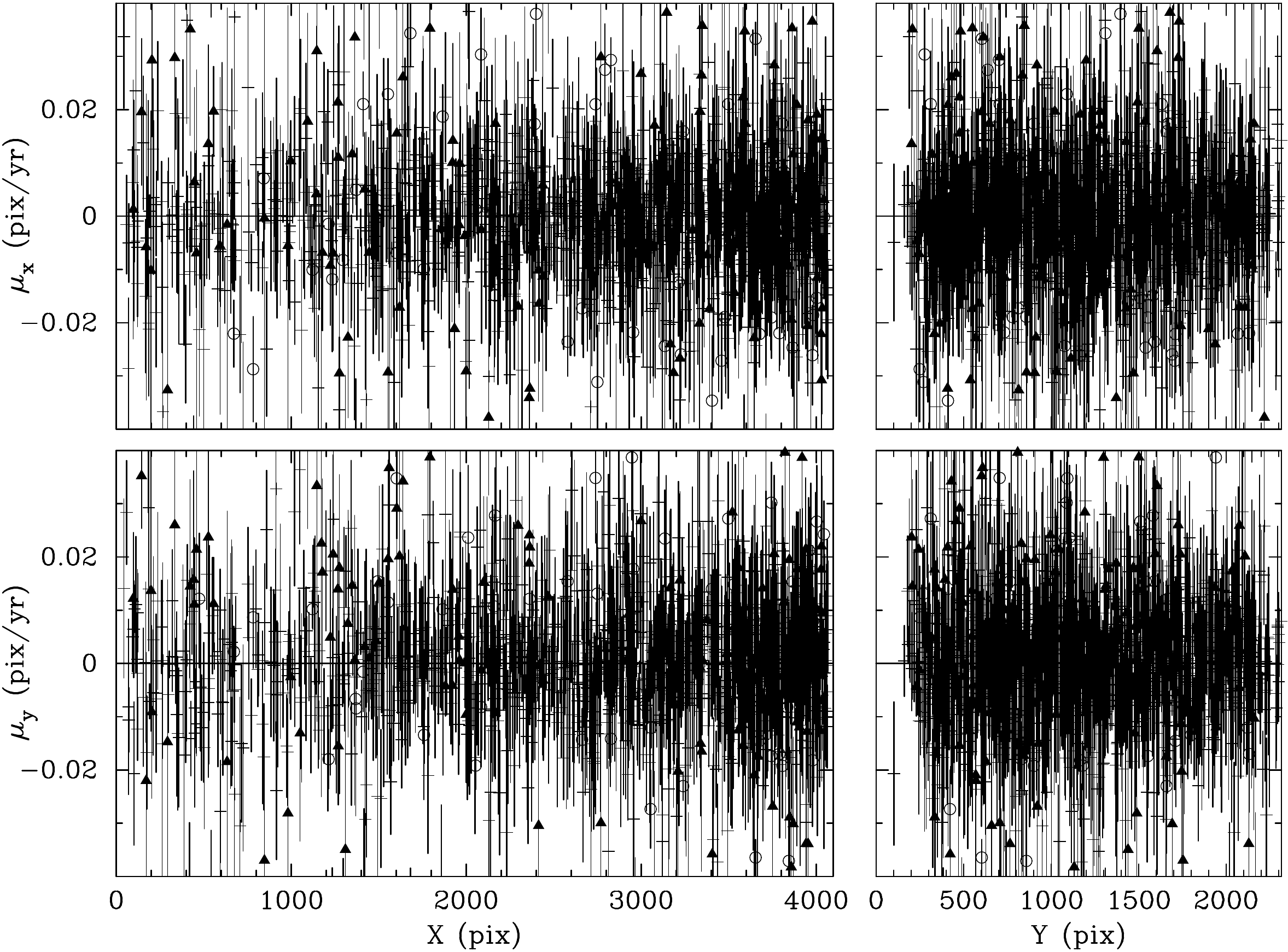}
\caption{The same as Figure~\ref{fig:mu_xy_leo21_s} for the Leo-II-2-UVIS2 field.}
\label{fig:mu_xy_leo22_t}
\end{figure}

   The next four figures (\ref{fig:mu_xy_leo21_s} ---
\ref{fig:mu_xy_leo22_t}) show $\mu_{x}$ as a function of $X$ (top
panels) and $\mu_{y}$ as a function of $Y$ (bottom panels) for the four
fields in Leo~II.  Here, the location of an object, $(X, Y)$, is the
location in the standard coordinate system.  
The likely stars of Leo~II are at rest in this system; thus, the
plus symbols representing them in the figures should show no trends with
$X$ or $Y$ and have $<\mu_{x}>=0$ and $<\mu_{y}>=0$.  A deviation from
this distribution would be indicative of errors in determining the
centroids.  A visual inspection of the four figures shows no significant
deviations, though the large scatter of the points may hide subtle
patterns. The field stars represented by open circles show, as expected,
a wide range of motions in the standard coordinate system owing to their
wide range of distances and velocities with respect to the Sun and no
dependence on $X$ or $Y$. The filled triangles representing the galaxies
should also show no trends, though their $<\mu_{x}>$ and $<\mu_{y}>$ are
expected not to be consistent with zero.  A visual inspection of the
figures indicates no significant trends with either $X$ or $Y$, but the
large scatter of the points may hide subtle patterns.

\begin{figure}[p]
\centering
\includegraphics[angle=-90,scale=0.60]{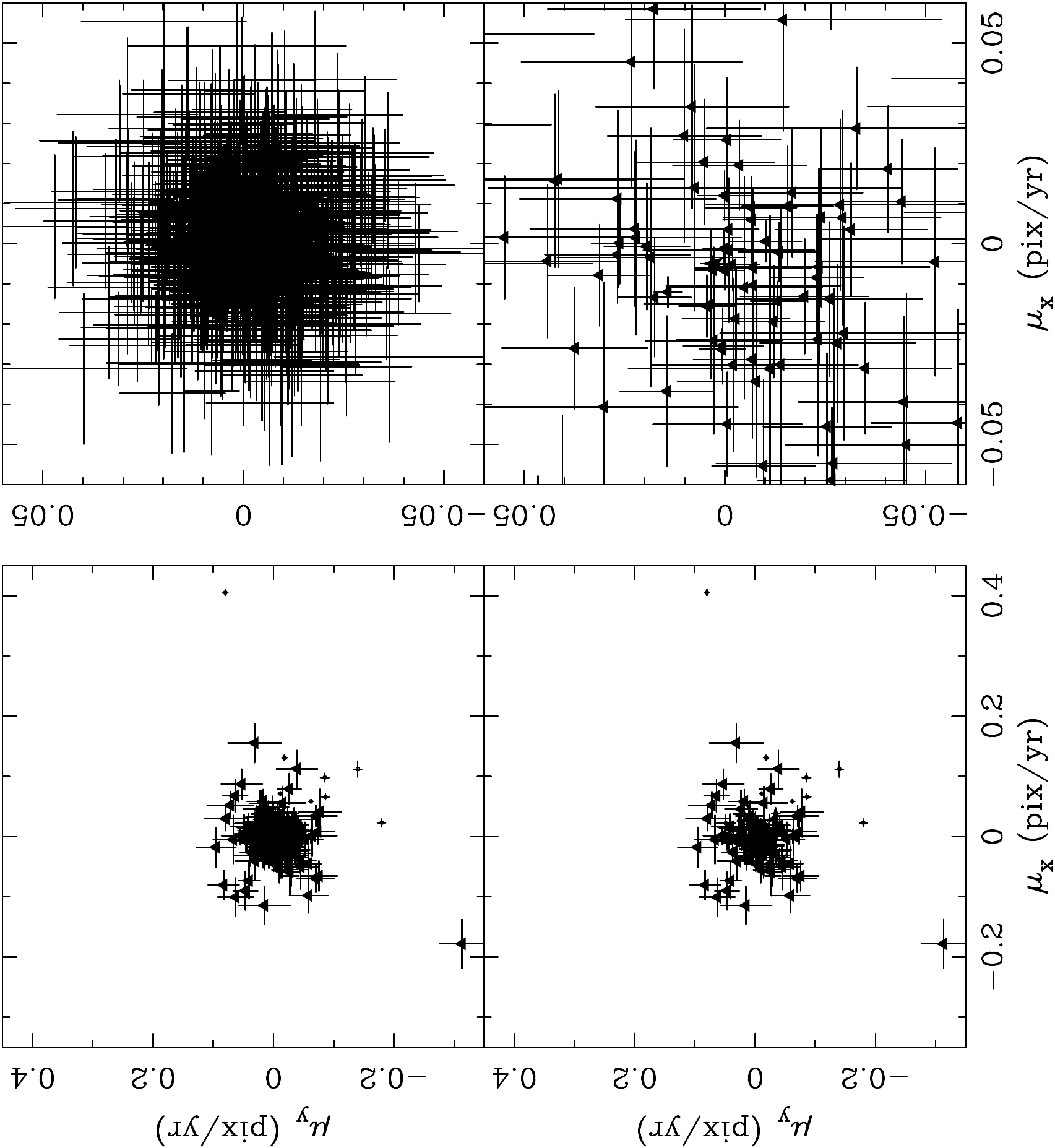}
\caption{Scatter plot of the measured values of $(\mu_{x},\mu_{y})$
for the objects in the Leo-II-1-UVIS1 field.  Top-left panel: the
plot for all of the objects. Bottom-left panel: that for the QSO,
galaxies, and field stars. Top-right panel: a zoomed-in view for the
likely stars of Leo~II.  Bottom-right panel: the same view as the above
panel for the QSO and galaxies.}
\label{fig:mux_muy_leo21_s}
\end{figure}

\begin{figure}[p]
\centering
\includegraphics[angle=-90,scale=0.60]{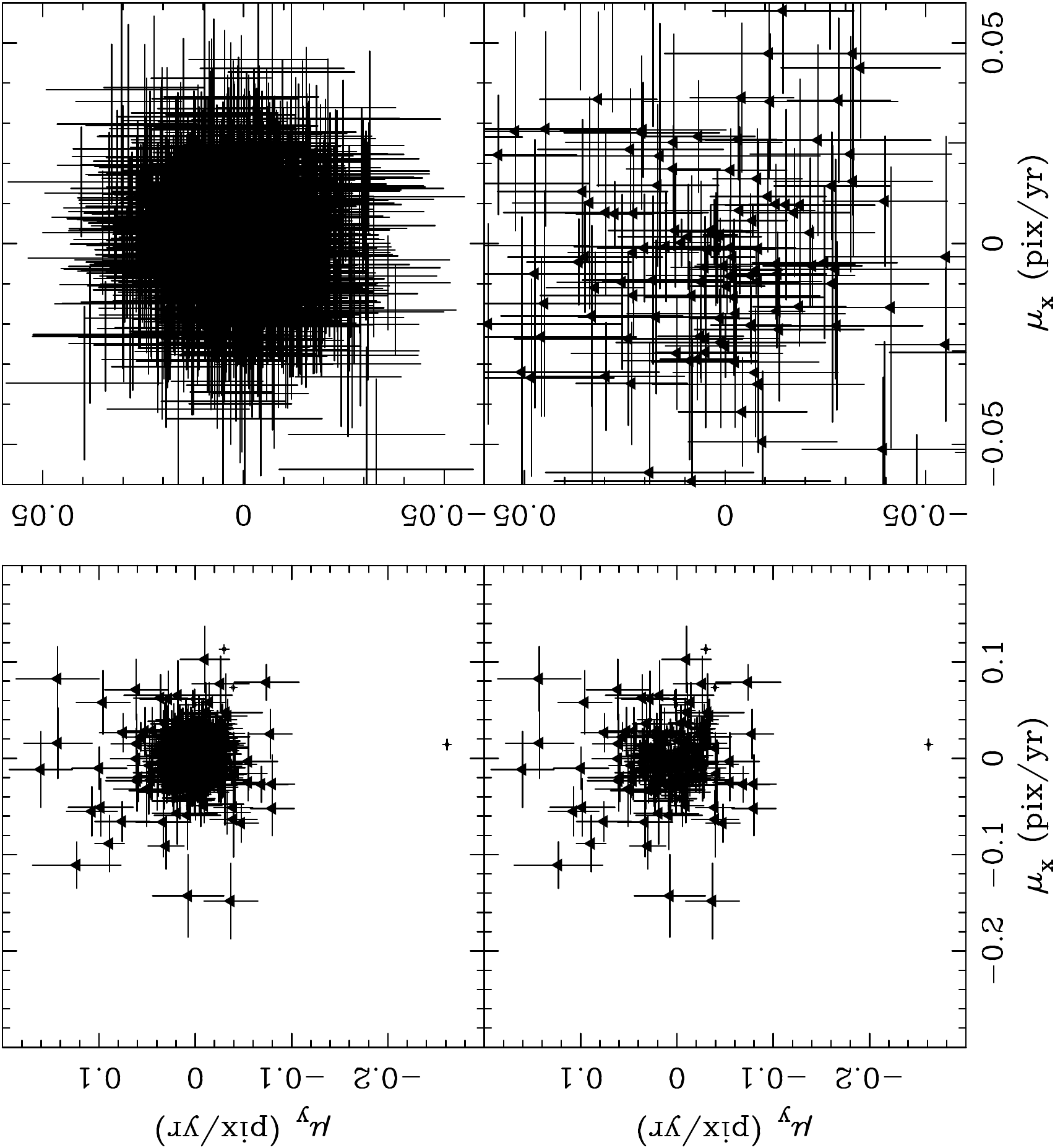}
\caption{The same as Figure~\ref{fig:mux_muy_leo21_s} for the Leo-II-1-UVIS2 field.}
\label{fig:mux_muy_leo21_t}
\end{figure}

\begin{figure}[p]
\centering
\includegraphics[angle=-90,scale=0.60]{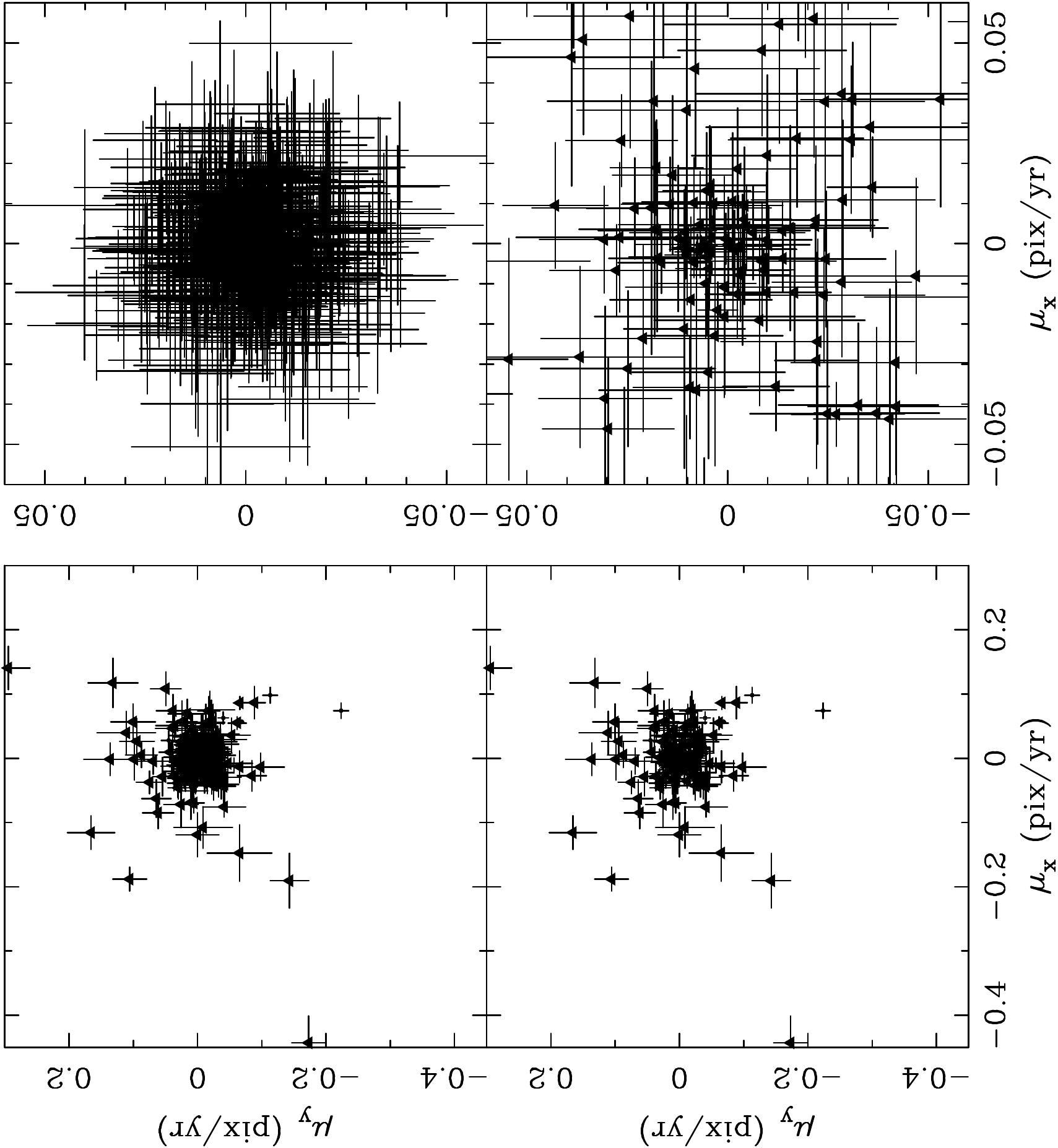}
\caption{The same as Figure~\ref{fig:mux_muy_leo21_s} for the Leo-II-2-UVIS1 field.}
\label{fig:mux_muy_leo22_s}
\end{figure}

\begin{figure}[p]
\centering
\includegraphics[angle=-90,scale=0.60]{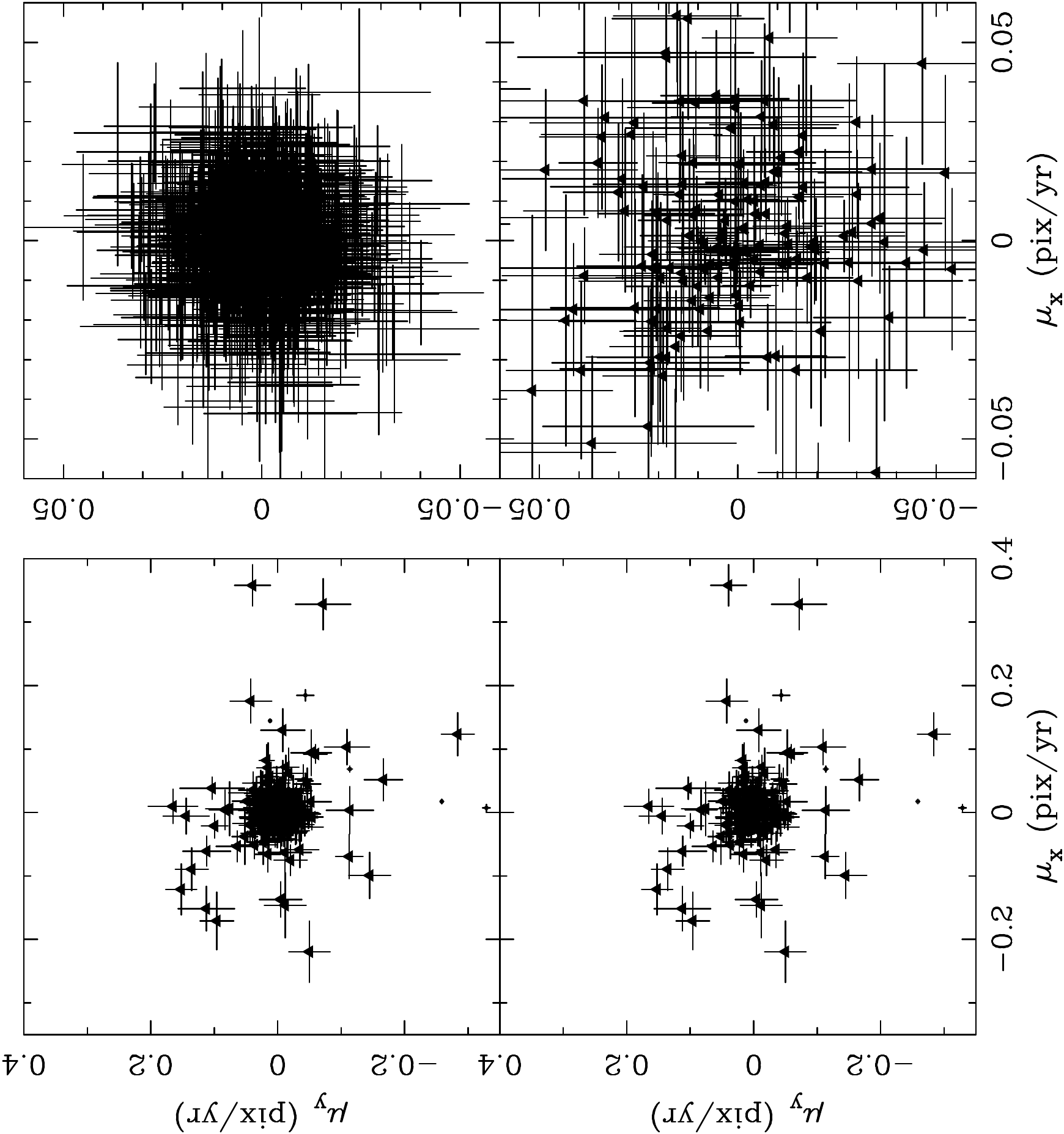}
\caption{The same as Figure~\ref{fig:mux_muy_leo21_s} for the Leo-II-2-UVIS2 field.}
\label{fig:mux_muy_leo22_t}
\end{figure}

       Figures~\ref{fig:mux_muy_leo21_s} --- \ref{fig:mux_muy_leo22_t}
are scatter plots of $(\mu_{x},\mu_{y})$ for the four fields.  The
top-left panel is for all of the objects; the meaning of the symbols is
the same as in the previous figures, with the one difference that the
size of the open circle representing a field star has been reduced to
enhance the visibility of the other, more important, objects. The scale
of the plot ensures that all of the objects with a measured motion are
depicted.  This plot should show the likely stars of Leo~II scattering
around $(0,0)$ with an approximately symmetric Gaussian distribution.
The galaxies should have a similar, though wider, distribution, but
around $(\mu_{x}, \mu_{y}) \neq (0,0)$. The plot in the lower-left panel
of the four figures is for the field stars and galaxies only; it has the
same scale as the panel above. The plot in the upper-right panel is for
the likely stars of Leo~II only, with a zoomed-in scale. The plot in the
lower-right panel has the same scale as the one directly above and is
for galaxies only.

    Visual inspection of the upper-right panels in
Figures~\ref{fig:mux_muy_leo21_s} -- \ref{fig:mux_muy_leo22_t} shows
that the distributions for the likely stars of Leo~II are well-behaved:
there are no obvious distortions from azimuthal symmetry. Similarly,
examining the lower-right panels shows that the galaxies are less
symmetrically distributed, though perhaps no more so than expected from
their smaller number and larger scatter.

Figure~\ref{fig:leo2} shows that the Leo-II-1-UVIS1 field overlaps a
portion of the Leo-II-2-UVIS1 field and that the Leo-II-1-UVIS2 field
overlaps portions of the Leo-II-2-UVIS1 and Leo-II-2-UVIS2 fields.
Because the orientations of the fields are the same to about 0\fdg 3,
the values of $\mu_x$ and $\mu_y$ derived from different fields can be
compared directly in the overlap regions.  Plotting the differences of
the $\mu_x$ values, $\Delta\mu_x$, or $\mu_y$ values, $\Delta\mu_y$,
\textit{versus} $X$, $Y$, or S/N shows no trends.  The weighted means
of $\Delta\mu_x$ and $\Delta\mu_y$ are zero within their uncertainties
for both stars and galaxies in all three overlap regions. The $\chi^2$
of the scatter around these means is approximately one per degree of
freedom for the galaxies when the motion uncertainties are increased as
is discussed in Section~\ref{sec:results}.  This argues that our final
proper motion uncertainties are correct.  The $\chi^2$ of the scatter
around the mean differences for the stars is larger than one per degree
of freedom, suggesting that the individual stellar motions have an
additional additive uncertainty of 0.002~pixel~yr$^{-1}$ ---
0.008~pixel~yr$^{-1}$ (8~mas~century$^{-1}$ --- 30~mas~century$^{-1}$),
depending on the region and component.  We have not incorporated this
additional uncertainty into our analysis because the uncertainty in the
transformation to the standard coordinate system, which is determined
collectively by the stars, contributes negligibly to the uncertainty for
all of the final measured proper motions presented in the next section.

After the completion of the analysis reported above, a software
correction for the decreasing CTE became available in the WFC3 reduction
pipeline \citep{ryan16}.  In response to a question from the referee, we
repeated the analysis using the corrected first-epoch images.  The
resulting changes in the X- and Y-components of the mean proper motion
of Leo~II were 0.1 and 0.6 of their uncertainties, respectively,
indicating that changes in the CTE have not significantly affected our
measured proper motion.  Moreover, examining the equivalents of the
left-hand panels of Figures~\ref{fig:mu_sn_leo21_s} ---
\ref{fig:mu_sn_leo22_t} clearly shows the stellar $\mu_y$ increasing
with increasing S/N in the UVIS1 data and the reverse trend in the UVIS2
data.  These trends are not present in our original figures.  A more
detailed examination of the data reveals that the trend is strongest for
stars with smaller Y-values in UVIS1 and with larger Y-values in UVIS2.
Thus, correcting for CTE losses is producing the expected pattern of
changes in the stellar $\mu_y$ values.  The surprise is that the
uncorrected data show no signature of CTE losses, while the corrected
data do.  Perhaps the pipeline software is over-correcting the losses
in the regime of our first-epoch data or perhaps the charge injection of
our second-epoch data is producing an effective CTE approximately equal
to that of our first epoch.  In any case, the effect of CTE losses on
our measured proper motion is small and we favor the value resulting
from our original analysis, which is presented below.

\section{RESULTS}
\label{sec:results}

Table~\ref{tab:pm} lists the six estimates of the proper motion in the
equatorial coordinate system.  In this article,
$(\mu_{\alpha},\mu_{\delta})$ are the components of the proper motion in
the direction of increasing right ascension and declination. Each
estimate with a ``g'' in column (2) is derived from the mean motion of
the galaxies in the standard coordinate system in the corresponding
field, whereas, each estimate with a ``q'' is derived from the motion of
the centrally-located QSO.  As in \citet{ppo15}, it is necessary to
increase the uncertainties in the galaxy motions by a multiplicative
factor to produce a $\chi^2$ per degree of freedom of one for the
average galaxy motion.  In the current work the factor is 1.78, 1.68,
2.27, and 1.95 for the fields Leo-II-1-UVIS1, Leo-II-1-UVIS2,
Leo-II-2-UVIS1, and Leo-II-2-UVIS2, respectively.  Since the QSOs were
treated as galaxies, all of the uncertainties in Table~\ref{tab:pm}
reflect these values.

\floattable
\begin{deluxetable}{lcrr}
\tablecolumns{4}
\tablewidth{0pt}
\tablecaption{Measured Proper Motion of Leo~II\label{tab:pm}}
\tablehead{
& &\colhead{$\mu_{\alpha}$}&\colhead{$\mu_{\delta}$}  \\
\colhead{Field}&\colhead{Ref.\tablenotemark{a}}&\multicolumn{2}{c}{(mas century$^{-1}$)}\\
\colhead{(1)}&\colhead{(2)}&\colhead{(3)}&\colhead{(4)}}
\startdata
Leo~II-1-UVIS1&g&$-22.7\pm 8.5$&$-12.5\pm8.7$\\
\noalign{\vspace{3pt}}
\multicolumn{1}{c}{$^{\prime\prime}$} &q&$-15.9\pm 22.7$&$-16.9\pm23.3$\\
\noalign{\vspace{3pt}}
Leo~II-1-UVIS2&g&$-10.1\pm7.3$&$-13.6\pm7.8$\\
\noalign{\vspace{3pt}}
Leo~II-2-UVIS1&g&$-1.5\pm9.0$&$-3.6\pm9.0$\\
\noalign{\vspace{3pt}}
\multicolumn{1}{c}{$^{\prime\prime}$} &q&$4.6\pm19.1$&$-22.3\pm28.3$\\
\noalign{\vspace{3pt}}
Leo~II-2-UVIS2&g&$1.9\pm6.7$&$-3.7\pm7.0$\\
\noalign{\vspace{1pt}}
\hline
\noalign{\vspace{1pt}}
Weighted mean&&$-6.9\pm3.7$&$-8.7\pm3.9$\\
\enddata
\tablenotetext{a}{Type of astrometric zero-point: g -- galaxy, q -- QSO.}
\end{deluxetable}

\begin{figure}[t!]
\centering
\includegraphics[angle=-90,scale=0.55]{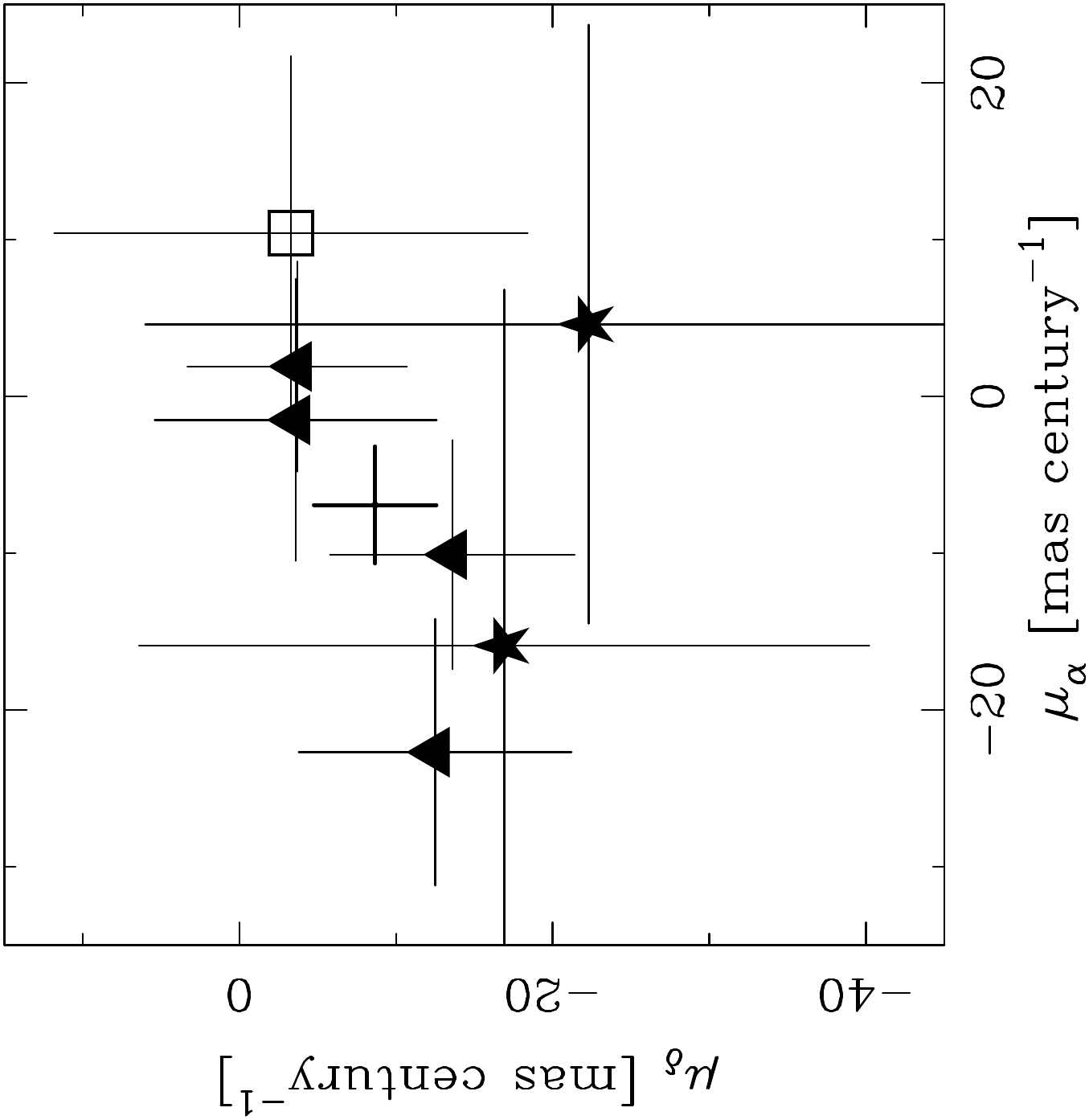}
\caption{The six independent estimates of the proper motion of Leo~II
from this study (filled triangles and stars) together with their weighted mean
(bold crossed error bars), and the measurement by \citet{lkrk11} (open
square).}
\label{fig:plotmu}
\end{figure}

Figure~\ref{fig:plotmu} shows the measurements from Table~\ref{tab:pm}
as filled triangles (galaxies) and filled stars (QSOs) together with
their weighted mean (bold crossed error bars) and the measurement by
\citet{lkrk11} (open square). The weighted mean is also given in the
bottom line of Table 3 and Equation~\ref{eq:epm} below.  This measured
proper motion of Leo~II is the main result of this study.
        
\begin{equation}
(\mu_{\alpha}, \mu_{\delta}) = (-6.9 \pm 3.7, -8.7 \pm 3.9)\
\textrm{mas}\ \textrm{century}^{-1}.
\label{eq:epm}
\end{equation}

Comparing the six measurements of $\mu_{\alpha}$ produces a
$\chi^{2}=6.3$ for five degrees of freedom, implying a probability $P =
0.28$ of seeing $\chi^{2}$ equal to or larger than this value. For the
corresponding measurements of $\mu_{\delta}$, $\chi^{2}=1.8$ and
$P=0.88$.  The agreement among the six measurements is acceptable;
therefore, the uncertainties in the measured proper motion are
realistic.

The measurement of the proper motion of Leo~II in this work is in
reasonable agreement with that from \citet{lkrk11}. The $\alpha$
components differ by 1.5~$\sigma$, while the $\delta$ components agree
within their uncertainties.  Our measurement has smaller uncertainties,
despite the seven times shorter time baseline, because the larger
passband and less-degraded CCDs produced a greater number of galaxies
determining the standard of rest.  The smaller pixel size and using a
template constructed for each galaxy also contributed to the better
astrometric precision.

         Equation~\ref{eq:gpm} expresses the measured proper motion of
Leo~II in the galactic coordinate system.

\begin{equation}
(\mu_{\ell}, \mu_{b}) = (6.2\pm 3.9, -9.2 \pm 3.7)\
\textrm{mas}\ \textrm{century}^{-1}.
\label{eq:gpm}
\end{equation}

The proper motion in Equations~\ref{eq:epm} or \ref{eq:gpm} can be
converted to the galactic-rest-frame (grf) proper motions
$(\mu_{\alpha}^{\mbox{\tiny grf}},\mu_{\delta}^{\mbox{\tiny grf}})$ and
$(\mu_{\ell}^{\mbox{\tiny grf}},\mu_{b}^{\mbox{\tiny grf}})$, which are
those seen by an observer at the location of the Sun and at rest with
respect to the Galactic center.  In addition, one can use the proper
motion to calculate the Galactocentric space velocity.  The components
in a cylindrical coordinate system are $\Pi$, $\Theta$, and $Z$, where
$\Pi$ is positive outwards, $\Theta$ is positive in the direction of
Galactic rotation, and $Z$ is positive in the direction of the north
Galactic pole.  The radial and tangential components in a spherical
coordinate system of the Galactocentric space velocity are $V_{r}$ and
$V_{t}$, respectively.  Calculating the above quantities requires
adopting specific values for the distance of the Sun from the Galactic
center, the circular speed of the LSR, the motion of the Sun with
respect to the LSR, and the distance and radial velocity of Leo~II with
respect to the Sun.  Table~\ref{tab:leoii} lists the adopted values and
Table~\ref{tab:derived} the derived motions.  The instantaneous space
velocity implies an orbital inclination of 68$^{\circ}$ with a 95\%
confidence interval of $(66^{\circ}, 80^{\circ})$.  The instantaneous
angular momentum vector points at $(\ell, b) = (192^{\circ},
-22^{\circ})$, with a 95\% confidence interval of $(149^{\circ},
237^{\circ})$ for $\ell$ and $(-24^{\circ}, -10^{\circ})$ for $b$.

\floattable
\begin{deluxetable}{lcc}
\tablecolumns{3}
\tablewidth{0pt} 
\tablecaption{Derived Motions of Leo~II\label{tab:derived}}
\tablehead{
\colhead{Quantity} &
\colhead{Value}    &
\colhead{Comment} \\
\colhead{(1)}&
\colhead{(2)}&
\colhead{(3)}}
\startdata
$(\mu_{\alpha},\mu_{\delta})^{\mbox{\tiny{Grf}}}$ &$(1.7 \pm 3.7, 11.0 \pm 3.9)$~mas~century$^{-1}$  & \tablenotemark{a} \\
$(\mu_{\ell},  \mu_{b})^     {\mbox{\tiny{Grf}}}$ &$(-10.0 \pm 3.9, 5.0 \pm 3.7)$~mas~century$^{-1}$  &  \tablenotemark{b} \\
$(\Pi, \Theta, Z)$                    &$(-37 \pm 38, 117 \pm 43, 40 \pm 16)$~km~s$^{-1}$&\tablenotemark{c} \\
$V_{r}$ & $21.9 \pm 1.5$~km~s$^{-1}$&\tablenotemark{d} \\
$V_{t}$ & $127 \pm 42$~km~s$^{-1}$& \tablenotemark{e} \\
$i$&$68^{\circ}$~$(66^{\circ}, 80^{\circ})$&\tablenotemark{f}\\
\enddata
\tablenotetext{a}{Galactic-rest-frame proper motion in the equatorial coordinate system.}
\tablenotetext{b}{Galactic-rest-frame proper motion in the galactic coorrdinate system.}
\tablenotetext{c}{Components of space velocity with respect to an observer at rest at the location of the LSR.}
\tablenotetext{d}{Radial velocity with respect to a stationary observer at the center of the Milky Way.}
\tablenotetext{e}{Tangential velocity with respect to a stationary observer at the center of the Milky Way.}
\tablenotetext{f}{Orbital inclination with respect to the Galactic plane.}
\end{deluxetable}

\section{DISCUSSION}
\label{sec:disc}

If Leo~II is bound gravitationally to the Milky Way, its Galactocentric
velocity provides a lower limit on the mass of the Galaxy.  If the
gravitational potential of the Milky Way is spherically symmetric, then
the lower limit on its mass $M$ is
\begin{equation}
\label{eq:mg}
M = \frac{R(V_r^2+V_t^2)}{2G},
\end{equation}
where $R$ is the Galactocentric radius of the galaxy and $G$ is the
universal constant of gravity.  Substituting in the measured
velocities from Table~\ref{tab:derived} and $R = 236$~kpc gives
$M = (5\pm 3) \times 10^{11}$~$M_\odot$.  The stated uncertainty in this
lower limit of the mass comes from the uncertainty in the velocity of
Leo~II.  The location of Leo~II is at about the virial radius of the
Milky Way --- probably slightly outside $r_{200}$ and inside $r_{vir}$
\citep[see][for example, for definitions of these quantities]{vdm2012}.
\citet{bhg16} reviews the observations and concludes that the virial
mass of the Milky Way is $M_{vir} = (1.3\pm 0.3) \times 10^{12}\
M_\odot$ and $r_{vir} = 282\pm 30$~kpc, which implies a mass inside the
location of Leo~II of $(1.2\pm 0.3) \times 10^{12}\ M_\odot$.  The above
lower limit on this mass is a factor of two below the range from
observations and implies that Leo~II is likely to be gravitationally
bound to the Galaxy and does not provide a strong constraint on its
total mass.

\citet{tkbi16} announced the discovery of the nearby dwarf galaxy
Crater~2 and points out that this galaxy is aligned in three dimensions
with the globular cluster Crater and the dwarf galaxies Leo~IV, Leo~V,
and Leo~II.  The study argues that these objects were accreted as a
bound group which was then tidally disrupted.  If this is the case, then
the Galactic-rest-frame proper motion of Leo~II should be at least
approximately aligned along the great circle passing through these
objects.  The great circle connecting Leo~II and Crater~2, the two ends
of the line of objects, has a position angle at the location of Leo II
of $167\fdg 1$ (measured from north, through east).  The equatorial
galactic-rest-frame proper motion from Table~\ref{tab:derived} has a
position angle of $9^\circ\pm 19^\circ$, which is $22^\circ$ from the
great circle and, thus, is aligned within the uncertainty.  The
direction of motion indicates that Leo~II would be leading the group.
\citet{tkbi16} notes that the galactocentric radial velocities of the
objects show that they are not on a single orbit.  Instead, the orbits
are co-planar with a range of energies probably acquired during the
disruption of the group at perigalacticon. Thus, all of the orbits
should have similar perigalacticons and the perigalacticon of Leo~II
should be at least as small as the smallest current Galactocentric
distance of the members of the group --- 120~kpc for Crater~2
\citep{tkbi16}. Integrating the orbit of Leo~II in a fixed Galactic
potential\footnote{The potential is that of the three-component mass
model of \citet{kslbh14} with the properties of the halo component
changed to match the virial mass and radius and the concentration
suggested by \citet{bhg16}.  The disk mass was also adjusted to keep the
circular velocity at the location of the Sun unchanged.  Actually using
the model of \citet{kslbh14}, which has a more centrally-concentrated
halo with a mass near the lower limit of the range found by
\citet{bhg16}, results in a perigalacticon for Leo~II of 204~kpc and a
23\% chance of having a perigalacticon of 120~kpc or less. Thus, our
conclusions are not strongly affected by the uncertainties in the mass
of the Galaxy.} starting with the space velocity in
Table~\ref{tab:derived} finds a larger perigalacticon of 161~kpc.
However, the 95\% confidence interval for this quantity is large, $(36,
233)$~kpc, and there is a 31\% chance that the perigalacticon is 120~kpc
or less.  Thus, the measured proper motion of Leo~II is consistent with
the proposal that it was part of a small group of objects which fell
into the Galaxy and was tidally disrupted.

\citet{pk13} discuss the membership of the satellites of the Milky Way
in the vast polar structure (VPOS).  The pole of the orbit of Leo~II
given at the end of Section~\ref{sec:results} allows a test of whether
its orbit is aligned with the plane of the VPOS.  The pole of the VPOS-3
plane is $(\ell,b) = (169\fdg 5, -2\fdg 8)$.  The two poles are
separated by about $29^\circ$, which is within the scatter of orbital
poles discussed by \citet{pk13}.

\section{SUMMARY}
\label{sec:sum}

We have measured the proper motion of the Leo~II dwarf galaxy using
images from four fields obtained with HST and WFC3 at two epochs
separated by approximately two years. The measurement uses compact
background galaxies and QSOs as the standard of rest.

The weighted proper motion of Leo~II is $(\mu_{\alpha},
\mu_{\delta})=(-6.9 \pm 3.7, -8.7 \pm 3.9)$~mas~century$^{-1}$.
The measurement from this study is in good agreement
with that from \citet{lkrk11} and has about a three times smaller
uncertainty in each component.

The method of using deep multi-epoch (preferably more than two) HST
imaging with compact background galaxies as the standard of rest is the
best approach today for measuring the proper motions of dwarf galaxies
in the halo of the Milky Way.  These observations are building up a
three-dimensional picture of the halo kinematics.  This picture contains
clues to help understand the formation of our Galaxy and its halo,
galactic interactions, galactic associations (streams or planar
alignments), and the amount and distribution of dark matter.

\acknowledgments

We thank Vera Kozhurina-Platais for providing us with easily readable
tables and polynomial coefficients for the geometrical distortions for
WFC3 with the F350LP filter.  We also thank Marla Geha and Joshua Simon
for their contributions in preparing the HST proposal that is
responsible for these observations. CP and SP acknowledge the financial
support of the Space Telescope Science Institute through the grant
HST-GO-11697.  EWO acknowledges support from the Space Telescope Science
Institute through the grant HST-GO-11697 and from the National Science
Foundation through the grant AST-1313006.

\facility{HST(WFC3), MMT(Blue channel spectrograph)}

\end{document}